\documentclass{ws-rv961x669}
\usepackage[square]{ws-rv-van}     
\usepackage{ws-rv-thm}     
\usepackage{subfigure}     
\usepackage{slashed}
\usepackage{bm}
\usepackage{hyperref}
\makeindex
\def\cropnote{\relax}
\usepackage{ulem}
\usepackage[dvipsnames]{xcolor}

\usepackage{verbatim}

\begin{document}

\chapter*{Resonant enhancement in leptogenesis}\label{reslg}

\author[]{P. S. B. Dev$^\ast$, M. Garny$^\dagger$\footnote{Corresponding Author.}, J. Klaric$^\dagger$,\\ P. Millington$^\ddagger$ and D. Teresi$^\S$}

\address{
$^\ast$ Department of Physics and McDonnell Center for the Space Sciences,
Washington University, St. Louis, MO 63130, USA
\\[3pt]
$^\dagger$ Physik Department T31, Technische Universit\"{a}t M\"{u}nchen,\\ James-Franck-Stra\ss e 1, D-85748 Garching, Germany
\\[3pt]
$^\ddagger$School of Physics and Astronomy, University of Nottingham,\\ Nottingham NG7 2RD, United Kingdom
\\[3pt]
$^\S$Service de Physique Th\'{e}orique, Universit\'{e} Libre de Bruxelles,\\[-2pt] Boulevard du Triomphe, CP225, 1050 Brussels, Belgium
\\[3pt]
$^1$ mathias.garny@tum.de
}

\begin{abstract}
{\bfseries Abstract:} Vanilla leptogenesis within the type I seesaw framework requires the mass scale of the right-handed neutrinos
to be above $10^9$\,GeV. This lower bound can be avoided if at least two of the sterile states are almost
mass degenerate, which leads to an enhancement of the decay asymmetry. Leptogenesis models that can be
tested in current and upcoming experiments often rely on this resonant enhancement, and
a systematic and consistent description is therefore necessary for phenomenological applications. In this
review article, we give an overview of different methods that have been used to study the saturation
of the resonant enhancement when the mass difference becomes comparable to the characteristic width of the Majorana
neutrinos. In this limit, coherent flavor transitions start to play a decisive role, and off-diagonal
correlations in flavor space have to be taken into account. We compare
various formalisms that have been used to describe the resonant regime and discuss under which circumstances
the resonant enhancement can be captured by simplified expressions for the CP asymmetry. Finally, we briefly
review some of the phenomenological aspects of resonant leptogenesis.
\end{abstract}


\body

\newpage

\tableofcontents

\section{Introduction}

The origin of the baryon asymmetry of the Universe poses one of the major challenges to our current understanding of particle physics
and therefore motivates a detailed investigation of baryogenesis in the context of extensions of the Standard Model (SM).
Leptogenesis~\cite{Fukugita:1986hr} provides a minimal realization, and the neutrino mass differences inferred from
observations of neutrino oscillations support the possibility of generating a lepton asymmetry via the
out-of-equilibrium decays of some heavy messenger particles in the early Universe.

Within the type I seesaw mechanism for generating the light neutrino masses~\cite{Mohapatra:1979ia, GellMann:1980vs, Yanagida:1979as, Minkowski:1977sc}, with Lagrangian
\begin{equation}\label{eq:L}
{\cal L}\ =\ {\cal L}_{SM} \:+\: \bar N_i \, i\slashed{\partial} \, N_i \:-\: \left(\frac12 \left(M_N\right)_{ij}\bar{N}_i^c N_j + \lambda_{\alpha i}\bar\ell_\alpha \phi^c N_i + \mbox{h.c.}\right)\,,
\end{equation}
it is well known that, for a hierarchical mass spectrum, the mass scale of the singlet neutrino states $N_i=P_R N_i$ should be
above around $\sim 10^9$\,GeV~\cite{Davidson:2002qv, Buchmuller:2002rq, Hamaguchi:2001gw, Buchmuller:2004nz, Giudice:2003jh} to account for the observed baryon asymmetry via out-of-equilibrium decays.
Here, $\ell_\alpha$ and $\phi$ are the $SU(2)_L$ lepton (with $\alpha=e,\mu,\tau$) and Higgs doublets, respectively, and the superscript $c$ denotes charge conjugation. The requirement on the heavy neutrino mass scale can be significantly relaxed if (at least)
two of the mass eigenvalues $M_i$ (say for $i=1,2$) are quasi-degenerate, i.e.~$\Delta M=M_2-M_1 \ll \bar M=(M_1+M_2)/2$.
In this case, the CP-violating decay asymmetry (summed over active flavors $\alpha$)
\begin{equation}
  \label{eq:treeCPparam}
  \epsilon_i \ =\ \frac{1}{(\lambda^\dag\lambda)_{ii}}\, \frac{\mbox{Im}[(\lambda^\dag\lambda)_{21}^2]}{8\pi}\frac{M_1M_2}{M_2^2-M_1^2}\;,
\end{equation}
originating from the interference of tree and wave (or self-energy) one-loop contributions --- the so-called $\varepsilon$-type or \emph{indirect} CP violation --- is resonantly enhanced~\cite{Liu:1993tg,Flanz:1994yx,Flanz:1996fb,Covi:1996fm,Covi:1996wh,Pilaftsis:1997jf,Buchmuller:1997yu}
and dominates over the contribution from vertex-corrections --- the so-called $\varepsilon'$-type or \emph{direct} CP violation.

Models of resonant leptogenesis (RL), within the framework of the type I seesaw model~\cite{Mohapatra:1979ia, GellMann:1980vs, Yanagida:1979as, Minkowski:1977sc}, supersymmetric extensions or in the context of the type~II~\cite{Mohapatra:1980yp,Magg:1980ut,Schechter:1980gr,Lazarides:1980nt,Wetterich:1981bx} and 
III~\cite{Foot:1988aq,Ma:1998dn} seesaw mechanisms,  provide the intriguing possibility to lower the mass scale to a realm that is accessible to current
and future experiments, either directly in collider searches, or indirectly, e.g.~via signatures of lepton-flavor-violating (LFV) processes.
Since the question of the origin of the baryon asymmetry is one of the key hints of physics beyond the SM, it is important to identify viable
scenarios that can be tested (or falsified) given the experimental efforts to explore the TeV scale physics (see the companion Chapter~\cite{leptogenesis:A05})\footnote{Seesaw physics below the Davidson-Ibarra bound is also motivated from Higgs naturalness arguments, which suggest the seesaw scale to be below $\sim 10^7$ GeV~\cite{Vissani:1997ys, Clarke:2015gwa, Bambhaniya:2016rbb}.}.

In this review, we give an overview of the current status of the theoretical description of resonant enhancement in leptogenesis, with special emphasis
on its saturation, which is expected to occur when the mass difference $\Delta M$ approaches the characteristic decay width of the Majorana neutrinos. In addition, we
discuss various RL scenarios that have been proposed. Whilst our discussions are far from being exhaustive, we intend to illustrate a number of possibilities that naturally explain the small mass
splitting required for resonant enhancement.

The saturation of resonant enhancement is commonly described by a modified decay asymmetry (cf.~Eq.~\eqref{eq:treeCPparam})
\begin{equation}\label{eq:cpA}
  \epsilon_i \ =\ \frac{1}{(\lambda^\dag\lambda)_{ii}}\, \frac{\mbox{Im}[(\lambda^\dag\lambda)_{21}^2]}{8\pi}\frac{M_1M_2(M_2^2-M_1^2)}{(M_2^2-M_1^2)^2+A^2}\,,
\end{equation}
which encompasses a ``regulator'' $A$ that controls the behavior of the decay asymmetry in the limit $\Delta M\to 0$. However, this modification is not sufficient to account for the dynamics that determines the final asymmetry
when saturation becomes relevant. Instead, it becomes necessary to take coherent flavor transitions among the Majorana neutrinos into account, which requires us to go beyond the usual description
based on (semi-classical) Boltzmann equations for the individual number densities. The description of the saturation regime shares some similarities to the description of leptogenesis via oscillations (see the companion Chapter~\cite{leptogenesis:A02}), although the mass scale is rather different and both are distinct scenarios. 

After providing some remarks on the history
of RL in Sec.~\ref{sec:earlyhistory}, we briefly review field-theoretic approaches that have been used to capture the saturation regime
in Sec.~\ref{sec:methodsoverview}. These are all based on the closed-time-path (CTP) description but employ different levels of approximations. 
Note that it is	straightforward to extend the field-theoretic description to cases where it is necessary to discriminate between 
different active flavors (cf.~e.g.~Eq.~\eqref{eq:epsilonalphabetalatetime}).
In Sec.~\ref{sec:weak}, we consider a simplified setup that allows us to compare the so-called two-time and Wigner-space approaches, and comment on the applicability of the usual Boltzmann description. Section~\ref{sec:scalar} discusses the so-called two-momentum/interaction-picture approach, applied to a scalar toy model, and its relation to the two-time description.
In Sec.~\ref{sec:strong}, we discuss an effective description of resonant enhancement applicable in the strong washout regime.
After this, we provide an overview of selected models of RL and the relevant phenomenological aspects in Sec.~\ref{sec:model}. Our conclusions are given in Sec.~\ref{sec:conclusions}.

\section{An early history of resonant leptogenesis}
\label{sec:earlyhistory}

In this section, we provide a very brief history of resonant enhancement of CP violation in leptogenesis up to the advent of the field-theoretic treatments that form the main focus of the present discussion. This overview is in no way complete. It is intended only to signpost some of the key steps in the early literature, and much of what follows draws heavily on the reviews that have punctuated the literature on this topic~\cite{Pilaftsis:1998pd,Buchmuller:2005eh,Mohapatra:2005wg,Davidson:2008bu,Pilaftsis:2009pk,Blanchet:2012bk,Fong:2013wr}.

The resonant enhancement of CP violation was first discussed in the context of the neutral kaon system (see Refs.~\cite{Pilaftsis:1997jf, Pilaftsis:1998pd}, and references therein). Here, the condition for resonant enhancement is satisfied through the slight breaking of the mass degeneracy of the $K^0$ and $\bar{K}^0$ states due to the weak interactions. In fact, it is from the literature on the kaon system that the terminology of $\varepsilon$- and $\varepsilon'$-type CP violation was adopted.

The potential for the resonant enhancement of CP-violating processes in the context of the baryon asymmetry was first acknowledged by Kuzmin, Rubakov and Shaposhnikov~\cite{Kuzmin:1985mm}, and the importance of the CP-conserving and CP-violating phases introduced by the wave-function renormalization of heavy, unstable particles was first emphasized in the work of Liu~\cite{Liu:1993bv}. The latter unpublished work was expanded upon by Liu and Segr\`{e}~\cite{Liu:1993tg}, who identified the possibility for resonant enhancement of the $\varepsilon$-type CP violation through the mixing of quasi-degenerate unstable states. Influential works on this kind of CP violation in the context of leptogenesis were then contributed by Franz, Paschos, Sarkar and Weiss~\cite{Flanz:1996fb};\footnote{Whilst the correct formula for the CP asymmetry appears in the second erratum to an earlier article by Flanz, Paschos and Sarkar~\cite{Flanz:1994yx}, it is in Ref.~\cite{Flanz:1996fb} that the physical importance of this result for the resonant enhancement of the CP-violation is acknowledged.} Covi, Roulet~\cite{Covi:1996fm} and Covi, Roulet and Vissani~\cite{Covi:1996wh}; Pilaftsis~\cite{Pilaftsis:1997dr,Pilaftsis:1997jf}
and Buchm\"uller and Pl\"umacher \cite{Buchmuller:1997yu}. 

In the degenerate limit, finite-order perturbation theory breaks down, since the wave-function amplitude for the CP-asymmetric mixing of the particles is inversely proportional to their mass splitting. Being unstable, these particles cannot appear as asymptotic states in $S$-matrix amplitudes, and their properties have instead been defined by the $S$-matrix elements of their stable, daughter particles~\cite{Veltman:1963th}. A consistent resummation scheme is needed in order to treat resonant transitions that involve unstable intermediate states, if their energy difference is comparable or smaller than their width. In Ref.\,\cite{Covi:1996fm}, an effective Hamiltonian approach
was used for a scalar toy model, similar to the description of $K^0-\bar K^0$ mixing.
The approach developed in Refs.~\cite{Papavassiliou:1995fq,Papavassiliou:1995gs,Papavassiliou:1996zn,Pilaftsis:1997dr,Pilaftsis:1997jf} (for a more comprehensive review of these considerations, see Ref.~\cite{Pilaftsis:1998pd} and references therein) is based on the so-called  pinch technique~\cite{Cornwall:1981zr,Cornwall:1989gv} and preserves important quantum field-theoretic properties such as CPT invariance and unitarity. 
By considering resonant contributions to two-to-two scattering amplitudes, it was also shown in Refs.~\cite{Roulet:1997xa, Flanz:1998kr} that $\varepsilon$-type CP violation 
is consistent with CPT and unitarity requirements, and that the deviation from thermal equilibrium is crucial to allow for self-energy type contributions
to the lepton asymmetry.
These findings were confirmed in Ref.~\cite{Buchmuller:1997yu} and extended to include a resummation of 
self-energy insertions.  

Eventually, all approaches lead to qualitatively similar results, namely
the conclusion that the regulator $A$ in Eq.~\eqref{eq:cpA} is of the order of the decay widths $\Gamma_i$ of the Majorana neutrinos (times the mass),
and all results agree quantitatively as long as the resonant enhancement 
is not yet within the saturation regime, i.e. for  $\Delta M\gg \Gamma_i$ \cite{Anisimov:2005hr,Dev:2014laa}.

It was quickly realized that resonant enhancement of the CP asymmetry allows for the heavy-neutrino mass scale to be lowered to the $\mathrm{TeV}$ scale~\cite{Pilaftsis:1997jf}, potentially allowing for observable signatures at colliders. This was illustrated comprehensively in the works by Pilaftsis and Underwood~\cite{Pilaftsis:2003gt,Pilaftsis:2005rv} to which the scenarios of \emph{resonant leptogenesis} owe their name. A number of phenomenological models that realize RL will be discussed in Sec.~\ref{sec:model}.

In early works on RL, the final asymmetry was obtained by solving systems of Boltzmann equations, supplemented with appropriately-resummed transition probabilities, as described above. In such semi-classical approaches, care must be taken to avoid double counting of decay and inverse decay processes due to two-to-two scatterings mediated by heavy neutrinos that go on resonance in the $s$ channel.
This requires a procedure of real intermediate state (RIS) subtraction (see Ref.~\cite{Kolb:1979qa} and the discussion in the companion Chapter~\cite{leptogenesis:A04}), which can be implemented by studying the analytic properties of the resonant $L$-violating scattering amplitudes.
It was these technical issues, in part, together with certain questions related to flavor effects (see the companion Chapter~\cite{leptogenesis:A01}), as well as the saturation of resonant enhancement, that motivated a move to approaches that allow one to obtain field-theoretic analogues of the Boltzmann equation from first principles. It is to these approaches that we now turn our attention.

\section{Overview of different methods}\label{sec:methodsoverview}

The resonant enhancement of the lepton asymmetry produced in the out-of-equilibrium decays of right-handed (RH) neutrinos has been described in a  number
of different theoretical frameworks. In this section, we briefly outline three approaches based on nonequilibrium Quantum Field Theory techniques \cite{Schwinger:1960qe,Keldysh:1964ud,Baym:1961zz,KadanoffBaym,Danielewicz:1982kk,Niemi:1983nf,Landsman:1986uw,Calzetta:1986cq,Knoll:2001jx,Blaizot:2001nr,CalzettaHu,Berges:2015kfa}
that have been employed more recently and that capture coherence
effects. Similar techniques have, for example, been used before in the context of electroweak baryogenesis \cite{Kainulainen:2001cn,Prokopec:2003pj,Konstandin:2004gy,Cirigliano:2009yt},
and are, from a broader perspective, used within a wide range of applications, such as for the description of the early stages of heavy ion collisions \cite{Blaizot:2001nr,Berges:2015kfa},  and in nuclear, atomic and condensed matter physics, see e.g. \cite{Danielewicz:1982kk,Bonitz,Leeuwen,Berges:2015kfa,pngf6} (see also Refs.~\cite{Buchmuller:2000nd,DeSimone:2007edo,DeSimone:2007gkc,Garny:2009rv,Garny:2009qn,Anisimov:2010aq,Beneke:2010wd} for early applications to leptogenesis). 

In general, coherence effects are  relevant for the saturation of the resonant enhancement for a very small splitting $\Delta M = M_2-M_1$ of the
mass eigenvalues of the order of the interaction rates $\Gamma$.  It is therefore important to ensure that these effects are fully captured in the systems of quantum Boltzmann equations describing the generation of the asymmetry.  This can be achieved either by employing the first-principles field-theoretic techniques that we discuss in this section or by careful treatment of so-called density matrix formalisms derived from the Liouville-von Neumann and Heisenberg equations (see the accompanying Chapter~\cite{leptogenesis:A01}).

All the methods outlined below are based on the so-called closed-time-path formalism (also known as Schwinger-Keldysh~\cite{Schwinger:1960qe,Keldysh:1964ud} or in-in formalism) and differ in the way in which various approximations and limits are implemented.
Before discussing the three different approaches in detail, we start by reviewing the common features and basic elements of the CTP framework.
In contrast to the more familiar in-out approach that is used to compute $S$-matrix elements, the CTP formalism can be used to compute
the time-evolution of expectation values. For example, the lepton number density (in the left-handed sector) can be related to the expectation value of the zeroth component of the lepton charge current via
\begin{equation}
  n_L(t)\ =\ \frac{1}{V}\int_V {\rm d}^3\mathbf{x}\; \langle J_{\ell}^0(t,\mathbf{x}) \rangle\;, \qquad {\rm with}\quad J_{\ell}^\mu \ =\ \sum_{\alpha\,=\,e,\mu,\tau} \bar\ell_\alpha\gamma^\mu\ell_\alpha\;.
\end{equation}
Whilst we should also consider the lepton number density stored in the right-handed charged leptons, it is sufficient to consider only the left-handed charged leptons when analyzing the source term for the asymmetry, as we do here.
For a treatment taking effects from active lepton flavor into account, it is also useful to define the matrix-valued
generalization
\begin{equation}
  n_{L\alpha\beta}(t)\ \equiv\ \frac{1}{V}\int_V {\rm d}^3\mathbf{x}\; \langle \bar\ell_\beta\gamma^{0}\ell_\alpha \rangle\;,
\end{equation}
from which the total asymmetry $n_L$ follows by taking the trace in flavor space.\footnote{For a small asymmetry, this definition coincides with the one
of Ref.~\cite{Beneke:2010dz}, $n_{L\alpha\beta}=q_{\ell\alpha\beta}$.}

\begin{figure}[!t]
\begin{center}
\includegraphics[width=0.5\textwidth]{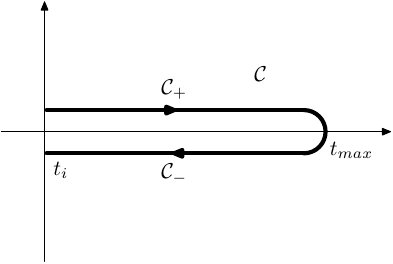}
\end{center}
\caption{\label{fig:CTP}The closed-time-path ${\cal C}$, starting at some initial time $t_i$, running (just above) the positive real time axis
to a maximal time $t_{max}>x^0,y^0$ and then running back to the initial time (just under) the positive real time axis. The two branches ${\cal C}_+$ and ${\cal C}_-$ are also known as the chronological and anti-chronological branches, respectively.}
\end{figure}

The expectation value of an operator $O(x)$ with respect to an initial state $|i\rangle$ is given in the interaction picture by
\begin{equation}
\langle O(x) \rangle\ =\ \langle i| U(t_i,x^0) O_I(x) U(x^0,t_i) | i \rangle \;,
\end{equation}
where $t_i$ is the initial time (relative to which the interaction picture is defined) and 
\begin{equation}
  U(t_1,t_2)\ =\ \left\{\begin{array}{ll}
    \mathrm{T} \exp \left( -\,i \int_{t_2}^{t_1} {\rm d}^4x\; {\cal L}_I \right)\;, &\qquad t_1>t_2 \\[0.5em] 
    \bar{\mathrm{T}} \exp \left(+\,i \int_{t_1}^{t_2} {\rm d}^4x\; {\cal L}_I \right)\;, &\qquad  t_2>t_1 
  \end{array} \right.
\end{equation}
is the time-evolution operator. Here, $\mathrm{T}$ is the usual time-ordering operator (later times to the left),
and $\bar{\mathrm{T}}$ is the anti-chronological time-ordering (later times to the right). The resulting time-ordering  
can be conveniently expressed by analytically continuing to the complex-time plane and introducing a time path $\mathcal{C}$ that starts at $t_i$,
runs to some time $t_{max}$ larger than $x^0$ and then back to $t_i$ (see Fig.\,\ref{fig:CTP}). This path can be parametrized by a function
 $t(u)$, where $u$ increases monotonically along the path. The time-ordering
in the expectation value can then be expressed in the compact form
\begin{equation}
\langle O(x) \rangle\ =\ \langle i| \mathrm{T}_{\cal C} \exp \left( - \, i \int_{\cal C} {\rm d}^4x\;{\cal L}_I \right) O_I(x) | i \rangle\;,
\end{equation}
where $\mathrm{T}_{\cal C}$ denotes path-ordering (i.e.~fields with larger values of $u$ to the left) and the time integration in the exponential is
along the path ${\cal C}$~\cite{Danielewicz:1982kk}. The expression for the expectation value then formally resembles the one for in-out matrix elements, with the important difference that the
usual time integration along the real axis is replaced by an integration along the CTP contour ${\cal C}$. It is also possible to consider mixed initial states specified by
an initial density matrix $\rho$ by replacing $\langle i|O| i \rangle \to \mbox{tr}(\rho O)$. 

The CTP formalism allows one to translate the path-integral techniques of the in-out formalism to the computation of expectation values. In particular, it can be used to generate a diagrammatic expansion for time-ordered $n$-point functions.
For example, the two-point function of the Higgs, lepton and Majorana neutrino fields are
\begin{eqnarray}
  \Delta_\phi^{ab}(x, y) \ &=&\  \langle \mathrm{T}_{\cal C} \phi^a(x)\phi^{*b}(y)\rangle\ \to\ \delta^{ab}\Delta_\phi(x, y)\;, \\
  S_{\ell\alpha\beta}^{ab}(x, y) \ &=& \ \langle \mathrm{T}_{\cal C} \ell_\alpha^a(x)\bar\ell^{b}_\beta(y)\rangle\ \to\ \delta^{ab}S_{\ell\alpha\beta}(x, y)\;, \\
  S_{Nij}(x, y) \ &=& \ \langle \mathrm{T}_{\cal C} N_i(x)\bar N_j(y)\rangle\;, 
\end{eqnarray}
where $a, b$ are $SU(2)$ indices and the arrows correspond to the two-point functions in an $SU(2)$ symmetric state, as appropriate far above the
electroweak scale. The time arguments $x^0$ and $y^0$ are attached to the CTP contour. If both time arguments are within the chronological branch ${\cal C}_+$, the time-ordering reduces to the usual one; if both are on the anti-chronological branch ${\cal C}_-$, one obtains an anti-time ordered two-point
function. If $x^0$ is on the anti-chronological branch and $y^0$ is on the chronological branch, one obtains the Wightman function $[S^>_{Nij}(x,y)]_{\lambda\kappa}=\langle N_{i\lambda}(x)
\bar N_{j\kappa}(y)\rangle$. Instead, if $y^0$ is on the anti-chronological branch and $x^0$ is on the chronological branch, one obtains $[S^{<}_{Nij}(x,y)]_{\lambda\kappa}=-\,\langle \bar N_{j\kappa}(y)N_{i \lambda}(x)\rangle$, where we note the minus sign that occurs in the definition for fermions. (We have displayed the Dirac indices $\lambda,\kappa$ for clarity.) It is also useful to define the statistical propagator $S^F$ and
the spectral function $S^\rho$ via
\begin{equation}\label{eq:SFrho}
  S_{Nij}(x, y) \ =\ S^F_{Nij}(x, y)\: - \: \frac{i}{2}\,\mbox{sgn}_{\cal C}(x^0-y^0)S^\rho_{Nij}(x, y)\,,
\end{equation}
where $\mbox{sgn}_{\cal C}$ is the signum function with respect to the path parameter $u$ that describes the position on the CTP contour, i.e.~$x^0=x^0(u)$.
The statistical propagator and spectral function are related to the Wightman functions via
\begin{equation}\label{eq:Sgtrless}
  S^\gtrless(x,y) \ =\ S^F(x,y)\: \mp\: \frac{i}{2}\,S^\rho(x,y)\,.
\end{equation}
Further frequently-used Green functions are the retarded, advanced and `hermitian' two-point functions
\begin{subequations}
\label{eq:SRAh}
\begin{eqnarray}
  S^R_{Nij}(x,y)\ &=&\ \theta(x^0-y^0)S_{Nij}^\rho(x,y) \,, \\
  S^A_{Nij}(x,y)\ &=&\ -\,\theta(y^0-x^0)S_{Nij}^\rho(x,y) \,, \\
  S^{\mathcal{H}}_{Nij}(x,y)\ &=&\ -\,\frac12\,\mbox{sgn}(x^0-y^0)S_{Nij}^\rho(x,y) \,,
\end{eqnarray}
\end{subequations}
where the Heaviside and signum functions take arguments on the real time axis.
Analogous definitions apply for the lepton and Higgs two-point functions, as well as the corresponding self-energies.
The spectral function and self-energy are also often denoted by ${\cal A}_N\equiv -\,i S_N^\rho/2$ and $ \Sigma_N^{\cal A}\equiv -\,i \Sigma_{\rho}/2$.

The lepton asymmetry is given by
\begin{equation}\label{eq:nL0}
  n_{L\alpha\beta}(t) \ =\  -\,\frac{g_w}{V}\int_V  {\rm d}^3\mathbf{x}\; \mbox{tr}[\gamma^0 S^{F}_{\ell\alpha\beta}(x, x)]\,,
\end{equation}
where $g_w=2$ counts the degenerate $SU(2)$ degrees of freedom.\footnote{In order to arrive at Eq.~\eqref{eq:nL0}, we have made the replacement \smash{$S_{\ell\alpha\beta}\to S_{\ell\alpha\beta}^F$}, discarding a potentially infinite but universal vacuum (zero-point) contribution arising from \smash{$S_{\ell\alpha\beta}^{\rho}$}. The latter can be avoided, and this replacement justified, by defining the signum function such that it vanishes for zero argument. Alternatively, the particle number density can be defined systematically by isolating the positive- and negative-frequency components of the Wightman functions, see e.g.~\cite{Beneke:2010wd, Millington:2012pf}. 
} Within the closed-time-path approach, the time-evolution of the lepton number can therefore be obtained from the
equation of motion of the two-point functions, given by the CTP Schwinger-Dyson equation. For example, for the Majorana neutrinos, it reads
\begin{equation}\label{eq:SD}
  i [S_{Nij}(x, y)]^{-1} \ =\  (i\slashed{\partial}_x - M_i\delta_{ij})\delta_{\cal C}(x^0-y^0)\delta^{(3)}(\mathbf{x}-\mathbf{y})\:-\: i\Sigma_{Nij}(x,y)\,,
\end{equation}
where $\Sigma_{Nij}$ is the (time-ordered) self-energy, which is a matrix in flavor space. The first term on the right-hand side is the inverse of the free two-point function, where $\delta_{\cal C}(x^0-y^0)$ is the Dirac delta function on the CTP contour: $\delta_{\cal C}(x^0-y^0)=\pm\,\delta(x^0-y^0)$ if $x^0$ and $y^0$ are both on $\cal C_{\pm}$ and $\delta_{\cal C}(x^0-y^0)=0$ otherwise.
The equations for the lepton and Higgs fields have a similar structure and can be derived, for example, from the two-particle-irreducible (2PI) effective action~\cite{Cornwall:1974vz}.
In vacuum or in thermal equilibrium, the two-point functions and self-energies depend
only on the difference of coordinates $x-y$, and the equation could be solved in Fourier space as a simple algebraic equation. Out of equilibrium,
space-time translational invariance is broken (by the boundary condition at $t_i$ and the expanding background), and two-point functions depend, in general, on
both arguments separately. For a spatially homogeneous system, which is of interest here,  spatial translational invariance implies a separate dependence only on the two time arguments,
and it is sometimes convenient to switch to Fourier space for the spatial momentum, using the mixed, \emph{two-time} representation $S_{Nij}(x^0, y^0,\mathbf{k})=\int\!\frac{{\rm d}^3\mathbf{k}}{(2\pi)^3}\,e^{i\mathbf{k}\cdot(\mathbf{x}-\mathbf{y})} S_{Nij}(x, y)$.
 In essence, the three methods described below differ in how the time-dependence is taken into account.

\subsection{Two-time formulation}

Within the two-time formulation, the equation of motion for the two-point functions $S_{Nij}(x^0, y^0,\mathbf{k})$ are solved taking into account the dependence on both
time arguments separately. This approach allows, in principle, for arbitrary deviations from thermal equilibrium and does not rely on specific approximations, in particular
for the spectral function and its relation to the statistical propagator. On the other hand, it is typically possible to solve the resulting equations only either numerically
or in specific simplified limits. Therefore, this approach has often been used as a benchmark that can be used to cross-check and assess the validity of various approximations~\cite{Berges:2015kfa}. It has been used to check the validity of the Boltzmann approach in the hierarchical case in Ref.~\cite{Anisimov:2010dk} and to assess the saturation of the resonant enhancement in Refs.~\cite{Garny:2011hg, Iso:2013lba}. The equation of motion can be brought to an explicit form by convoluting the Schwinger-Dyson equation~\eqref{eq:SD} with the two-point function and inserting the decomposition~\eqref{eq:SFrho}, along with a similar decomposition of the self-energy. In this way, one obtains
\begin{eqnarray}\label{eq:KB}
  (i\slashed{\partial} - M_i)S^F_{Nij}(x,y) \ &=&\ \int_{t_i}^{x^0} {\rm d}^4z\;\Sigma^\rho_{Nik}(x,z)S^F_{Nkj}(z,y) \nonumber\\&&\qquad-\: \int_{t_i}^{y^0} {\rm d}^4z\;\Sigma^F_{Nik}(x,z)S^\rho_{Nkj}(z,y)\,,\nonumber\\
  (i\slashed{\partial} - M_i)S^\rho_{Nij}(x,y)\ &=&\ \int_{y^0}^{x^0} {\rm d}^4z\; \Sigma^\rho_{Nik}(x,z)S^\rho_{Nkj}(z,y) \,.
\end{eqnarray}
The self-energy of the RH neutrino states is given at one-loop order by
\begin{eqnarray}\label{eq:SigmaN}
  \Sigma_{Nij}(x,y) \ &=&\ -\,g_w\lambda_{\alpha i}^*\lambda_{\beta j} P_L S_{\ell\alpha\beta}(x,y)\Delta_\phi(x,y)P_R \nonumber\\
  && \qquad - \: g_w \lambda_{\alpha i}\lambda_{\beta j}^* P P_L \bar S_{\ell\alpha\beta}(\bar x,\bar y)\bar \Delta_\phi(\bar x,\bar y)P_RP\,,
\end{eqnarray}
where $P_R$ and $P_L$ are the right and left chiral projection operators, and $\bar S_\ell( x,y) \equiv CP S_\ell( \bar y,\bar x)^{\mathsf{T}} (CP)^{-1}$ and $\bar\Delta_\phi(x,y)=\Delta_\phi( \bar y,\bar x)$ are CP-conjugate
lepton- and Higgs two-point functions with $\bar x=(x^0,-\,\mathbf{x})$, $P=\gamma_0$ and $C=i\gamma_2\gamma_0$. Analogous equations
can be derived for the lepton and Higgs fields, and the lepton self-energy at ${\cal O}(\lambda^2)$ is given by
\begin{equation}\label{eq:SigmaEll}
  \Sigma_{\ell\alpha\beta}(x,y)\ =\ -\,\lambda_{\alpha i}\lambda_{\beta j}^* P_R S_{Nij}(x,y)P_L\Delta_\phi(y,x)\;.
\end{equation}
Taking the time derivative of Eq.~\eqref{eq:nL0} and using the equation of motion for the lepton two-point function yields an evolution
equation for the total lepton asymmetry:
\begin{eqnarray}\label{eq:nL}
  n_{L\alpha\beta}'  &=& ig_w\!\int\!\frac{{\rm d}^3\mathbf{p}}{(2\pi)^3} \int_{t_i}^t\!{\rm d}t' \, \mbox{tr}\Big(\Sigma^\rho_{\ell\alpha\gamma}(t,t',\mathbf{p})S^F_{\ell\gamma\beta}(t',t,\mathbf{p}) \,-\, \Sigma^F_{\ell\alpha\gamma}(t,t',\mathbf{p})S^\rho_{\ell\gamma\beta}(t',t,\mathbf{p})\nonumber\\
  && {} -\,S^\rho_{\ell\alpha\gamma}(t,t',\mathbf{p})\Sigma^F_{\ell\gamma\beta}(t',t,\mathbf{p})\, +\, S^F_{\ell\alpha\gamma}(t,t',\mathbf{p})\Sigma^\rho_{\ell\gamma\beta}(t',t,\mathbf{p}) \Big)\,,
\end{eqnarray}
where we have used the Fourier representation with respect to the spatial coordinates and the trace is over the Dirac indices.
The expansion of the Universe can be taken into account in a simple way~\cite{Beneke:2010wd} by moving to conformally flat coordinates, wherein the metric $g_{\mu\nu}=a\eta_{\mu\nu}$ is proportional to the Minkowski metric $\eta_{\mu\nu}$ up to the scale factor $a$. 
In this case, $t\equiv \eta$ is the conformal time, which is related to the physical time via ${\rm d}t\equiv {\rm d}\eta={\rm d}t^{\rm phys}/a$,
and $p$ is the comoving momentum, with $p^{\rm phys}=p/a$. 
Apart from this reinterpretation of the time coordinates and momenta, the only additional change is that the mass term
is replaced by the time-dependent one $M_i \to aM_i$. The physical density of lepton number is related
to the comoving one by $n_L^{\rm phys}=n_L/a^3$, i.e.~$n_L'={\rm d}n_L/{\rm d}\eta=a^4({\rm d}n_L^{\rm phys}/{\rm d}t^{\rm phys}+3Hn_L^{\rm phys})$.

For a small asymmetry, the right-hand side of Eq.~\eqref{eq:nL} can be expanded in the CP-odd part of the
lepton and Higgs two-point functions, given by $\delta S_{\ell\alpha\beta}(x,y)=S_{\ell\alpha\beta}(x,y)-\bar S_{\ell\alpha\beta}(x,y)$ and $\delta\Delta_\phi(x,y)=\Delta_\phi(x,y)-\bar\Delta_\phi(x,y)$, respectively. The lowest order in this
expansion yields the source term $S$ for the lepton asymmetry, and the first order yields the washout term $W$, and we have
\begin{equation}
  n_{L\alpha\beta}'\ \simeq\ g_w S_{\alpha\beta} \:+\: W_{\alpha\beta}\;.
\end{equation}
We have taken out a factor of $g_w=2$ in the definition of the source term following the conventions of Ref.~\cite{Garbrecht:2014aga}.
 Since the lepton and Higgs fields are kept
close to thermal equilibrium by their gauge interactions, a useful approximation is to consider them as a thermal bath~\cite{Anisimov:2010dk}. For the source
term, this means one makes the following replacements at leading order:
\begin{subequations}
\begin{align}
\Delta^F(t,t',\mathbf{p})\ &\to \ \frac{1}{2p}\,[1+2/(e^{p/T}-1)]\cos[p(t-t')]\;,\\
\Delta^\rho(t,t',\mathbf{p})\ & \to \ \frac{1}{p}\,\sin[p(t-t')]\;,\\
S_{\ell\alpha\beta}(t,t',\mathbf{p})\ &\to\ \delta_{\alpha\beta}\,S_{\ell}(t,t',\mathbf{p})\;,
\end{align}
\end{subequations}
where $p\equiv|\mathbf{p}|$ and
\begin{subequations}
\begin{align}
S^F_{\ell}(t,t',\mathbf{p}) \ &\equiv\ \frac12[1-2/(e^{p/T}+1)]\Big(-\!\tfrac{\mathbf{p}\cdot\gamma}{p}\cos[p(t-t')]-i\gamma_0\sin[p(t-t')]\Big)\;,\\
S^\rho_{\ell}(t,t',\mathbf{p}) \ &\equiv\ -\tfrac{\mathbf{p}\cdot\gamma}{p}\sin[p(t-t')]+i\gamma_0\cos[p(t-t')]\;.
\end{align}
\end{subequations}

In this approximation, Eq.~(\ref{eq:KB}) yields a self-consistent equation for the non-equilibrium two-point function of the Majorana neutrinos,
which is a matrix in flavor space. Solving this equation and inserting the solution in the right-hand side of Eq.~(\ref{eq:nL}) then yields the source term that describes the generation of the asymmetry~\cite{Garny:2011hg}:
\begin{eqnarray}\label{eq:SourceTwoTime}
  S_{\alpha\beta} \ &=&\ i\lambda_{\alpha i}\lambda_{\beta j}^* \int\!\frac{{\rm d}^3\mathbf{p}}{(2\pi)^3} \int_{t_i}^{t} {\rm d}t'\; \mbox{tr}\Big[P_R \Delta S_{Nij}^\rho(t,t',\mathbf{p})
  P_L S_{\ell\phi}^F(t',t,\mathbf{p}) \nonumber\\
  && \qquad -\: P_R \Delta S_{Nij}^F(t,t',\mathbf{p})
  P_L S_{\ell\phi}^\rho(t',t,\mathbf{p}) \Big]\,,
\end{eqnarray}
where $\Delta S_{Nij}(t,t',\mathbf{p})=S_{Nij}(t,t',\mathbf{p})-\bar S_{Nji}(t,t',\mathbf{p})$ and $S_{\ell\phi}(x,y)=S_{\ell}(x,y)\,\Delta_\phi(x,y)$
is the lepton-Higgs loop.

The main ingredient needed to compute the lepton asymmetry is a solution of
the Kadanoff-Baym~\cite{Baym:1961zz,KadanoffBaym} equations (\ref{eq:KB}). The latter form a system of self-consistent equations for $S^{\rho}_{Nij}(t,t',\mathbf{p})$ and $S^{F}_{Nij}(t,t',\mathbf{p})$, i.e.~the solution
appears both on the right- and left-hand sides.
Diagrammatically, the
self-consistency can be
related to an iterative solution of the Schwinger-Dyson equation in an ouf-of-equilibrium situation, which resums insertions of lepton/Higgs loops.
This is precisely the type of diagram that is resonantly enhanced for a small mass splitting. The resummation ensures that the solution is valid
also when the mass-squared splitting is of the same order as the self-energy (which is related to the width via the optical theorem).
The full solution, which has a matrix structure in flavor space, can, in general, be obtained only numerically,
 although analytic and semi-analytic solutions can be found in specific limits. This approach therefore provides the possibility to assess
the validity of various approximation schemes.

\subsection{Wigner-space formulation}
\label{sec:wigner}

The Wigner-space approach to non-equilibrium Green's functions is widely used to describe transport phenomena~\cite{Prokopec:2003pj}.
It was first applied to RL for small mass splittings in Ref.~\cite{Garbrecht:2011aw} and further
developed, e.g., in Refs.~\cite{Iso:2014afa, Garbrecht:2014aga, Drewes:2016gmt}.
The main idea is to derive a generalization to the usual Boltzmann equations, starting from the Schwinger-Dyson
equation (\ref{eq:SD}), that is applicable if the two-point functions vary much more rapidly with respect to the relative
coordinate $r=x-y$ than the central coordinate $X=(x+y)/2$. This is generally expected close to thermal equilibrium. In order to exploit this separation of scales,
 one considers the two-point functions in Wigner space:
\begin{equation}
 i S_{Nij}(p,X)\ =\ \int {\rm d}^4r\;e^{ip\cdot r } S_{Nij}(X+r/2, X-r/2)\,.
\end{equation}
Analogous definitions apply for
the lepton and Higgs two-point functions, as well as the self-energies, and for the individual components.
Note that we have introduced a factor of $i$ on the left-hand side such that, in Wigner space, the conventions match precisely the ones in Refs.~\cite{Prokopec:2003pj, Beneke:2010wd}
(apart from the retarded and advanced functions, which differ by a sign).
For spatially homogeneous systems, the two-point functions depend only on the time coordinate $t\equiv X^0$, which characterizes the ``slow'' variation,
while the ``fast'' variation is described by $p^0$.
In order to derive equations of motion in Wigner space, one needs the Wigner transform of a convolution of two two-point functions:
\begin{equation}
  \int {\rm d}^4(x-y)\; e^{ip\cdot(x-y)} \int {\rm d}^4z\;A(x,z)B(z,y) \ =\  e^{-i\diamond}\{A(p,X)\}\{B(p,X)\}\;,
\end{equation}
where the operator $\diamond\{\cdot\}\{\cdot\}=\frac12 (\partial^{(1)}\cdot\partial_k^{(2)}-\partial_k^{(1)}\cdot\partial^{(2)})\{\cdot\}\{\cdot\}$ generates nested Poisson brackets, $\partial\equiv\partial_X$
and the superscripts refer to the first and second arguments, respectively. In Wigner space, the Schwinger-Dyson equations for the Wightman functions, obtained from Eq.~(\ref{eq:SD}), read~\cite{Konstandin:2004gy}
\begin{equation}
\label{eq:WignerSD}
\Big(\slashed{p}+\frac{i}{2}\slashed{\partial}- M\Big)S_N^{<,>}\:-\:e^{-i\diamond}\{\Sigma_N^{\mathcal{H}}\}\{S_N^{<,>}\}\:-\:e^{-i\diamond}\{\Sigma_N^{<,>}\}\{S_N^{\mathcal{H}}\}\ =\ {\cal C}_N\,,
\end{equation}
where ${\cal C}_N=\frac12 e^{-i\diamond} (\{\Sigma_N^{>}\}\{S_N^<\}-\{\Sigma_N^{<}\}\{S_N^>\})$ are the collision terms and we have left implicit a matrix notation for the flavor structure in which $M=M_i\delta_{ij}$. Equation~\eqref{eq:WignerSD} can also be obtained directly from the Kadanoff-Baym equations (\ref{eq:KB}) by rewriting the finite time integrals as $\int_{t_i}^{x^0}{\rm d}z^0(...)=\int_{t_i}^\infty {\rm d}z^0 \frac12[1+\mbox{sgn}(x^0-z^0)](...)$, before using Eq.~(\ref{eq:Sgtrless}) and Eq.~(\ref{eq:SRAh}) to express all two-point functions and self-energies in terms of $<$, $>$ and hermitian components, and finally performing a Wigner transformation in the limit $t_i\to-\infty$.
With the exception of this last step, both sets of equations are fully equivalent.
In Wigner space, the Kadanoff-Baym equations can be solved by truncating a gradient expansion,
which formally corresponds to an expansion in $\partial \ll p$. The convergence of this expansion relies on the separation of (time-)scales alluded to before. At lowest order in gradients, one keeps only the zeroth order, i.e.~$e^{-i\diamond}\to 1$, such that there are no Poisson brackets.

The $>$ and $<$ components of the self-energies at zeroth order in the gradient expansion, obtained from Wigner transforming Eqs.~(\ref{eq:SigmaN}) and (\ref{eq:SigmaEll}), are
given by
\begin{eqnarray}
  i\Sigma_{Nij}^{\gtrless}(k) \: &=&\: g_w \int \frac{{\rm d}^4k'}{(2\pi)^4}\frac{{\rm d}^4k''}{(2\pi)^4}(2\pi)^4\delta^{(4)}(k-k'-k'') \Big( \lambda^*_{\alpha i}\lambda_{\beta j} P_L iS_{\ell\alpha\beta}^\gtrless(k')P_R i\Delta^\gtrless_{\phi}(k'') \nonumber\\
 && \qquad+\: \lambda_{\alpha i}\lambda^*_{\beta j} C [P_L iS_{\ell\beta\alpha}^\lessgtr(-k')P_R ]^T C i\Delta^\lessgtr_{\phi}(-k'') \Big)\,, \\
 i\Sigma_{\ell\alpha\beta}^{\gtrless}(k) \: &=&\:  \lambda_{\alpha i}\lambda_{\beta j}^* \int \frac{{\rm d}^4k'}{(2\pi)^4}\frac{{\rm d}^4k''}{(2\pi)^4}(2\pi)^4\delta^{(4)}(k-k'-k'') 
  P_R i S_{Nij}^{\gtrless}(k')P_L i\Delta^{\lessgtr}_{\phi}(-k'') \nonumber\,,
\end{eqnarray}
where we have omitted the argument $X$ for brevity.
The equation of motion for the lepton asymmetry, again at zeroth order in the gradient expansion, is given by
\begin{eqnarray}
  n_{L\alpha\beta}' \ &=&\ g_w \int \frac{{\rm d}^4k}{(2\pi)^4}\;\frac12\, \mbox{tr} \Big[ i\Sigma^>_{\ell\alpha\gamma}(k) i S^<_{\ell\gamma\beta}(k) \:-\: i\Sigma^<_{\ell\alpha\gamma}(k) i S^>_{\ell\gamma\beta}(k)\nonumber\\
  && \qquad -\: iS^>_{\ell\alpha\gamma}(k) i \Sigma^<_{\ell\gamma\beta}(k) \:+\: iS^<_{\ell\alpha\gamma}(k) i \Sigma^>_{\ell\gamma\beta}(k) \Big]\;.
\end{eqnarray}
Similarly to the two-time formulation, we can extract the leading-order source term by replacing the lepton and Higgs two-point functions by the free, thermal
expressions~\cite{Garbrecht:2011aw}, which yields~\cite{Garbrecht:2014aga}
\begin{equation}
	\label{BG:src}
	S_{\alpha\beta} \ =\  - \,\lambda_{\alpha i}\lambda^*_{\beta j} \int \frac{{\rm d}^4k}{(2\pi)^4}\;\mbox{tr} \left[ P_R i\delta S_{Nij}(k)2P_L\slashed{\Sigma}_N^{\cal A}(k) \right]\,,
\end{equation}
where \smash{$\delta S_{Nij}(k)$} is the deviation of \smash{$S^F_{Nij}(k)$} from thermal equilibrium and \smash{$\slashed{\Sigma}_N^{\cal A}$} is the reduced self-energy, as given in Ref.~\cite{Drewes:2012ma}.
Note that $i\slashed{\Sigma}_N^{\cal A}$ coincides with the leading term in the gradient expansion of the Wigner transform of $-\,iS^\rho_{\ell\phi}(x,y)/2$.
This expression can also be obtained from the source term \eref{eq:SourceTwoTime} in the two-time approach by expanding to zeroth order in gradients (using $\bar S^\rho_{\ell\phi}(k)=S^\rho_{\ell\phi}(k)=S^\rho_{\ell\phi}(-k)$), taking $t_i\to-\infty$, and neglecting $\Delta S_{Nij}^\rho$ and $S_{\ell\phi}^{\mathcal{H}}$.
Therefore, within the approximations implicit to the truncation of the gradient expansion, these two formulations are consistent with one another.

As was the case for the two-time formulation, the equation for the asymmetry needs to be complemented by an evolution equation for the Majorana neutrino two-point functions.
It has been shown in Ref.~\cite{Garbrecht:2011aw} that, for a quasi-degenerate mass spectrum $\Delta M \ll M$, the equations for the two-point functions can be reduced to
kinetic equations for a matrix of RH neutrino distribution functions  $\delta f_{ahij}$ ($a=0,1,2,3$, $h=\pm$ is the helicity). Specifically, we can make the decomposition
\begin{align}
	\label{BG:2pf}
	\delta S_{N}(k) \ &=\  \sum_{h\,=\,\pm}\delta S_{Nh}(k)\,, \\
	-\,i\gamma^0 \delta S_{Nh}\ &=\ \frac14(1+h\hat k^i\sigma^i)\otimes \rho^a 2\pi\delta(k^2-a^2\bar M^2)2k^0 f_{ahij}\,,
\end{align}
where $\sigma^i$ and $\rho^a$ are Pauli matrices, $\hat k^i\equiv k^i/|{\bm k}|$ and $\bar M\equiv (M_1+M_2)/2$.
 As discussed in Ref.~\cite{Garbrecht:2011aw}, the resonant enhancement is well described to leading order in $\Delta M/\bar M$ and $\Gamma/\bar M$, even when taking the on-shell limit for the spectral function $\propto \delta(k^2-a^2\bar M^2)$, which leads to important simplifications.
 
 In the non-relativistic limit, it is sufficient to track $\delta f_{0hij}$ (see below). The resulting kinetic equation, when neglecting $S^{\mathcal{H}}$ and $\Sigma^{\mathcal{H}}$ (which describe a thermal
mass shift, see e.g.~Ref.~\cite{Drewes:2016gmt}), is given by
\begin{equation}\label{eq:df0hij}
  \delta f_{0h}'\:+\:\frac{i}{2k^0}\,\big[(aM)^2,\delta f_{0h}\big]\:+\:{f_{\rm eq}}'\:+\:\{\Gamma_h,\delta f_{0h}\}\ =\ 0\,,
\end{equation}
where expressions for $\Gamma_h$ and the source term expressed in terms of $\delta f_{0h}$
 can be found in Ref.~\cite{Garbrecht:2014aga}. This equation has the form of a density matrix equation in flavor space
for the quasi-degenerate pair of Majorana neutrinos with helicities $h=\pm$, wherein the commutator term describes oscillations. 

Substituting the two-point function~\eqref{BG:2pf} into the source term~\eqref{BG:src}, one obtains
\begin{align}
	S_{\alpha\beta} \ &=\ - \sum_{i,j} \lambda_{\alpha i}\lambda^*_{\beta j} \int \frac{{\rm d}^4k}{(2\pi)^4}\;\mbox{tr} \left[ P_R i\delta S_{Nij}(k)2P_L\slashed{\Sigma}_N^{\cal A}(k) \right]\,,\\
	&=\ \sum_{i,j} \sum_{h\,=\,\pm} \int \frac{{\rm d}^3\mathbf{k}}{(2\pi)^3}\; \lambda_{\alpha i}\lambda^*_{\beta j}
		\left\{
			\frac{k \cdot \hat{\Sigma}^\mathcal{A}_N(k)}{k_0}\,\big[\delta f_{0hij}(k)-\delta f_{0hij}^*(k)\big]
		\right.\\
	&\qquad \left.
	\left.
	+\:h\,\frac{\tilde{k} \cdot \hat{\Sigma}^\mathcal{A}_N(k)}{k_0}\,\big[\delta f_{0hij}(k)+\delta f_{0hij}^*(k)\big]
	\right\}
	\right|_{k^0\,=\,\omega(\mathbf{k})} \,,
\end{align}
where we have introduced $\tilde{k}\equiv (|\mathbf{k}|,k^0 \mathbf{k}/|\mathbf{k}|)$ and $\omega(\mathbf{k}) \equiv\sqrt{\mathbf{k}^2 + a^2 \bar{M}^2}$.

\subsubsection{Non-relativistic approximations}\label{sec:nrapp}

If the masses of the RH neutrinos are much larger than the temperature close to the time of freeze-out ($\bar M \gg T$),
the momentum modes that do not satisfy $|\mathbf{k}|\ll a \bar{M}$ are Maxwell suppressed. It is therefore sufficient to approximate the four-momenta by
\begin{align}
	k^\mu\ =\ (k^0,\mathbf{k})\ \approx\ (\pm\,a \bar M,\mathbf{0})\,,
\end{align}
The same reasoning allows us to neglect the thermal corrections to the self-energy (for discussions of thermal corrections to rate equations for leptogenesis, see  the accompanying Chapter~\cite{leptogenesis:A04}; see also Refs.~\cite{Drewes:2016gmt, Biondini:2015gyw}), i.e.
\begin{align}
	\left(\hat{\Sigma}^{\mathcal{A}}_N(k)\right)^\mu\ \approx\ \mathrm{sgn}(k^0)\,\frac{k^\mu}{32 \pi}\,,
\end{align}
whose contractions with the two four-vectors appearing in the evolution equations read as
\begin{align}
	k \cdot \hat{\Sigma}^\mathcal{A}_N(k) \ =\ \mathrm{sgn} (k^0)\, \frac{a^2 \bar M^2}{32 \pi}
	\,,&&
	\tilde{k} \cdot \hat{\Sigma}^\mathcal{A}_N(k) \ =\ 0\,.
\end{align}
Substituting this result into Eq.~\eqref{eq:df0hij}, we may integrate over the momentum modes to obtain
\begin{subequations}\label{eq:dn0hij}
\begin{align}
	\delta n^{\prime}_{0h}\: +\: \frac{a}{2 \bar{M}}\,i\left[ M^2, \delta n_{0h} \right]\: +\: n_{\rm eq}^{\prime}\ &=\ 
	-\,\frac{g_w a \bar{M} }{32 \pi}\left\{ {\rm Re}\, \lambda^\dagger \lambda , \delta n_{0h} \right\}\,,\\
	\delta \bar{n}^{\prime}_{0h}\: -\: \frac{a}{2 \bar{M}}\,i\left[ M^2, \delta \bar{n}_{0h} \right]\: +\: n_{\rm eq}^{\prime}\ &=\ 
	-\,\frac{g_w a \bar{M} }{32 \pi}\left\{ {\rm Re}\, \lambda^\dagger \lambda , \delta \bar{n}_{0h} \right\}\,,
\end{align}
\end{subequations}
where we have introduced the comoving non-equilibrium number densities of the sterile neutrinos
\begin{align}
	\delta n_{0h}\ =\ \int \frac{{\rm d}^3 \mathbf{k}}{(2\pi)^3}\;\delta f_{0h}(+\,\omega(\mathbf{k}),\mathbf{k})\,,\qquad \delta \bar{n}_{0h}\ =\ \int \frac{{\rm d}^3 \mathbf{k}}{(2\pi)^3}\;\delta f_{0h}(-\,\omega(\mathbf{k}),\mathbf{k})\,.
\end{align}
From the Majorana condition, we also obtain the relation between the positive- and negative-energy states $\delta n_{0h} = \delta \bar{n}_{0h}^{*}$.
Notice that the resulting comoving density and the equations of motion are helicity independent in this non-relativistic limit, i.e.~$\delta n_{0h} = \delta n_{0-h}$.
Inserting the approximated number densities and self-energies into the flavored source term yields
\begin{align}
	S_{\alpha \beta} \ &=\ \frac{a \bar{M}}{16 \pi} \sum_{i,j} \lambda_{\alpha i} \lambda^*_{\beta j}
	\left(\delta n_{0hij} - \delta \bar{n}_{0hij} \right)\,,
	\label{src:nrapp}
\end{align}
where the sum over helicities has been evaluated to give an overall factor of two. Furthermore, from the
Majorana constraint $\delta n_{0hij}=\delta \bar{n}^{*}_{0hij}$, one can see that only the off-diagonal
correlations enter the source term.

\subsection{Two-momentum formulation and interaction picture}

\subsubsection{Two-momentum formulation}

Rather than working with Wigner or time-domain functions, one can also work in a two-momentum representation of the non-equilibrium two-point functions~\cite{Herranen:2010mh,Fidler:2011yq, Herranen:2011zg, Millington:2012pf,Millington:2013isa}. We first note that the lepton asymmetry, as given in Eq.~\eqref{eq:nL0}, can be written in the following way:
\begin{equation}
n_{L\alpha\beta}(t)\ =\ -\,\frac{g_{w}}{V}\,\int_V{\rm d}^3\mathbf{x}\int{\rm d}^4y\,\delta^{(4)}(x-y)\,\mathrm{tr}\,[\gamma^0S^F_{\ell\alpha\beta}(x,y)]\;,
\end{equation}
where $x^0=t$. Making use of a double Fourier transformation, we insert
\begin{equation}
S^F_{\ell\alpha\beta}(x,y)\ =\ \int\!\frac{{\rm d}^4p}{(2\pi)^4}\int\!\frac{{\rm d}^4p'}{(2\pi)^4}\;e^{-ip\cdot x}\,e^{ip'\cdot y}S^F_{\ell\alpha\beta}(p,p')\;,
\end{equation}
where the sign convention on the four-momenta is chosen such that exact energy-momentum conservation corresponds to $p=p'$. After performing the coordinate integrals, we arrive at
\begin{equation}
\label{3_nLmomentum_PM}
n_{L\alpha\beta}(t)\ =\ -\,\frac{g_{w}}{V}\,\int\!\frac{{\rm d}^3\mathbf{p}}{(2\pi)^3}\int\frac{{\rm d}p_0}{2\pi}\int\frac{{\rm d}p_0'}{2\pi}\;e^{-i(p_0-p_0')t}\,\mathrm{tr}\,[\gamma^0 S^F_{\ell\alpha\beta}(p,p')]\;,
\end{equation}
with $\mathbf{p}'=\mathbf{p}$. Note that we have assumed the volume $V$ to be sufficiently large that we have approximate conservation of three-momentum. 

In order to see that the vacuum (zero-point) terms in $S^F_{\ell\alpha\beta}$ do not contribute to the final asymmetry, it is illustrative to restrict to a single flavor and consider the two-momentum representation of the tree-level equilibrium function (assuming a non-vanishing chemical potential). The latter is given by (see e.g.~Ref.~\cite{LeBellac})
\begin{align}
S_{\ell\alpha\beta}^{F,0}(p,p')\ &=\ \pi(\slashed{p}+m)\delta(p^2-m^2)(2\pi)^4\delta^{(4)}(p-p')\nonumber\\&\qquad \times\:\Big[\theta(p_0)\big(1-2f[E(\mathbf{p})]\big)\:+\:\theta(-p_0)\big(1-2\bar{f}[E(\mathbf{p})]\big)\Big]\;.
\end{align}
Inserting this into Eq.~\eqref{3_nLmomentum_PM}, the terms independent of the particle and anti-particle distributions $f$ and $\bar{f}$, viz.~the particle and anti-particle vacuum contributions, cancel after evaluation of the spinor trace:
\begin{align}
n_{L\alpha\beta}(t)\ &\supset\ -\,g_{w}\!\int\!\frac{{\rm d}^4p}{(2\pi)^3}\,\frac{1}{2}\,\delta(p^2-m^2)\,\mathrm{tr}\,[\gamma^0 (\slashed{p}+m)] \nonumber\\ &=\ -\,2\,g_{w}\!\int\!\frac{{\rm d}^4p}{(2\pi)^3}\,p_0\,\delta(p^2-m^2)\ =\ 0\;.
\end{align}
We then obtain the expected result
\begin{align}
\label{3_doublemomn_PM}
n_{L\alpha\beta}(t)\ &=\ g_{w}g_{s}\,\int\!\frac{{\rm d}^3\mathbf{p}}{(2\pi)^3}\;\Big[f[E(\mathbf{p})]\:-\:\bar{f}[E(\mathbf{p})]\Big]\;,
\end{align}
where we have made the identification $V\equiv(2\pi)^3\delta^{(3)}(\mathbf{0})$ and $g_s=2$ accounts for the sum over spin polarizations. 

The two-momentum representation is particularly advantageous in the case of particle mixing, and we will illustrate this explicitly in Sec.~\ref{sec:IPframework} and Sec.~\ref{sec:scalar} within the so-called interaction-picture framework of non-equilibrium field theory~\cite{Millington:2012pf,Millington:2013isa}.

\subsubsection{Interaction-picture framework}
\label{sec:IPframework}

The \emph{Fermi Golden Rule} tells us that the probability for a given process to take place is proportional to the period of time $t$ and the volume $V$ over which that process is permitted to occur. In the case of scattering probabilities, this four-volume factor is infinite,\footnote{More precisely, we assume a significant separation between the microscopic scales over which individual interactions take place and the macroscopic scales over which initial states are prepared, the interactions are turned on and the final state is collected. It is this separation of scales that allows us to assume that energy and momentum are exactly conserved in individual scattering events, when, in reality and by virtue of Heisenberg’s uncertainty principle, the former can be violated by an amount inversely proportional to the duration of our experiment and the latter by an amount inversely proportional to the volume of our experiment.} and it is captured by the product of identical energy-momentum-conserving Dirac delta functions that results from squaring the relevant matrix elements. We deal with this mathematically ill-defined product of distributions, which yield so-called \emph{pinch singularities}, by means of the \emph{Fermi Trick}: we replace one of the four-dimensional delta functions by a global factor of the four-volume $Vt$, which can be divided out in order to define the transition rate per unit volume. In this way, the convergence of our perturbation theory (modulo infra-red effects) is controlled by the interaction strength, $g$ say, and not the product of the interaction strength and the time over which those interactions are permitted, i.e.~$gt$.  The same is not true out-of-equilibrium: perturbation theory is controlled by both $g$ and the product $gt$.\footnote{The interpretation of pinch singularities in non-equilibrium perturbation theory in terms of the Fermi Golden Rule was identified concisely in Ref.~\cite{Greiner:1998ri}.}  For instance, consider the exponential approach to equilibrium. The deviation from equilibrium goes like $\delta f(t)\sim e^{-\Gamma t}\delta f(0)$, where the rate $\Gamma\propto g$. The exponential has a convergent Taylor series expansion only for $t<1/g$, and it would therefore appear fruitless to try to describe non-equilibrium processes perturbatively for $t>1/g$~\cite{Berges:2004yj}. In corollary, when we are in thermodynamic equilibrium, i.e.~$\delta f(t)=0$, our perturbation theory should again be controlled by $g$ only, and this is indeed what one finds. The pinch-singular terms are absent in thermodynamic equilibrium by virtue of the Kubo-Martin-Schwinger (KMS) condition or, equivalently, the fluctuation-dissipation theorem and detailed balance (see e.g.~Ref.~\cite{Landsman:1986uw}).

The interaction-picture approach~\cite{Millington:2012pf,Millington:2013isa} yields a perturbation theory built from two-point functions of the form
\begin{equation}
S^{0,ab}_{\ell\alpha\beta}(x,y,t)\ \equiv\ \mathrm{tr}\,\{\rho(t)\,{\rm T}_{\mathcal{C}}[\ell^a_{\alpha}(x)\,\bar{\ell}^b_{\beta}(y)]\}\;.
\end{equation}
Most importantly, the quantum statistical density operator $\rho(t)$ is regarded as an unknown with respect to the perturbation series, and its form is fixed self-consistently only after solving the system of Boltzmann-like equations derived from this perturbation series. In addition, integrals over intermediate interaction vertices are necessarily restricted to the finite time domain $[0,t]$.\footnote{Here, we assume that the interactions are turned on instantaneously, but this can easily be generalized to allow for smooth switching functions.} This results in the absence of exact energy-conserving Delta function; specifically~\cite{Millington:2012pf,Millington:2013isa},
\begin{equation}
\delta\Big(\sum E\Big)\ \to\ \delta_t\Big(\sum E\Big)\ \equiv\ \frac{t}{2\pi}\,\mathrm{sinc}\,\Big(\sum Et/2\Big)\;,
\end{equation}
where the function $\delta_t$ is analytic for all finite $t$. It is in this way that perturbation theory is viable. For early times, the absence of exact delta functions ensures that there can be no pinch singularities~\cite{Millington:2012pf,Millington:2013isa}.\footnote{For early discussions of the problem of pinch singularities in non-equilibrium field theory, see e.g.~Refs.~\cite{Altherr:1994jc,Weldon:1991ek,Altherr:1994fx,Dadic:1998yd,Bedaque:1994di}.} For intermediate times, the would-be pinch-singular terms grow as a power law in $t$ (as we would expect from the Fermi Golden Rule). Finally, for late times, the distribution functions themselves approach equilibrium exponentially, and the pinch singularities begin to cancel. The perturbation theory therefore remains under control for all times.

Given the canonical algebra of the theory, the tree-level propagators can be evaluated explicitly without the need for quasi-particle approximation. The tree-level propagators depend on a set of time-dependent distribution functions of the following form:
\begin{subequations}
\begin{align}
f_{\alpha\beta}^{ab}(\mathbf{p};s,\mathbf{p}';s';t)\ \equiv\ \mathrm{tr}\,\rho(t)\,b^{b\dag}_{\beta}(\mathbf{p}',s')b^a_{\alpha}(\mathbf{p},s)\;,\\
\bar{f}_{\alpha\beta}^{ab}(\mathbf{p};s,\mathbf{p}';s';t)\ \equiv\ \mathrm{tr}\,\rho(t)\,d_{\alpha}^{\dag}(\mathbf{p},s)d_{\beta}^b(\mathbf{p}',s')\;.
\end{align}
\end{subequations}
We restrict ourselves to considering only flavor, isospin and helicity correlations but, more generally, this set of distribution functions can readily be extended to keep track also of pair (particle-anti-particle) correlations, as well as correlations between different particle species. It is clear that the propagators can take a complicated form in generality, and complete expressions can be found in Ref.~\cite{Millington:2012pf} (for scalars), Refs.~\cite{Dev:2014wsa} (for scalars in the presence of particle mixing) and Ref.~\cite{Dev:2014laa} (for fermions in the presence of particle mixing).

Whilst various field-theoretic ingredients, derived by means of the interaction-picture approach, have been used in the construction of semi-classical rate equations for RL~\cite{Dev:2014laa,Dev:2014tpa,Dev:2015wpa}, this framework has, to date, been applied in full only to toy scalar models~\cite{Dev:2014wsa,Dev:2015dka,Kartavtsev:2015vto} (see also Sec.~\ref{sec:scalar}). Therein, it has been used to cross-check existing approximation schemes and make concrete comparisons between semi-classical and field-theoretic descriptions. In this context and in the weak washout regime, the interaction-picture approach has been shown to yield results identical to those obtained in the Heisenberg picture~\cite{Kartavtsev:2015vto}. Its particular advantage lies in the ability to identify concretely (i) the spectrum of states that are being counted by the number densities and (ii) the processes that are contributing to the evolution of these number densities. Loop-wise perturbative approximations to the former and latter are referred to, respectively, as \emph{spectral} and \emph{statistical truncations}. The independence of these two truncations means that the interaction-picture approach is much closer in spirit to semi-classical approaches, and it is for this reason that this framework has found such utility in making comparisons between existing semi-classical and field-theoretic results.

\section{Two-time versus Wigner and Boltzmann}\label{sec:weak}

In this section, we consider a simplified setup that allows us to compute the resonantly-enhanced asymmetry, and its saturation,
based on the first-principle approaches outlined above. The primary objective of this exercise is to scrutinize the resonant
enhancement mechanism and to identify suitable approximation schemes that can be applied also in realistic scenarios.
For that purpose, we generalize the setup considered in Ref.~\cite{Anisimov:2010dk} to the case of a quasi-degenerate mass spectrum
and study the asymmetry that is generated by the relaxation of the Majorana neutrino fields into thermal equilibrium.
Concretely, this means we disregard washout processes for the moment and adopt a static background described by a thermal
bath of lepton and Higgs fields at a constant temperature $T$.

This setup allows us to obtain a solution for the out-of-equilibrium Majorana neutrino two-point functions, including flavor off-diagonal
correlations, without further approximations and based on the two-time formulation~\cite{Garny:2011hg}.
This is the essential ingredient for computing the generation of the asymmetry. It takes into account all potential sources for a saturation
of the resonant enhancement, while being free of \emph{a priori} assumptions. In addition, following Ref.~\cite{Anisimov:2010dk}, one can also take into
account a finite width for the lepton and Higgs fields in a simplified manner, in order to investigate whether they contribute to the
saturation of the resonant enhancement. 
So as to provide a quantitative discussion, we assume for simplicity that the Majorana neutrinos are non-relativistic,
i.e.~that $T\ll M_i$.

\paragraph{Boltzmann approach:} For reference, we first quote the result for the total lepton asymmetry obtained via the conventional
Boltzmann approach. When applied to the simplified setup considered in this section, it is given by 
\begin{eqnarray}
  n_L(t) \ &=&\  \int\frac{{\rm d}^3\mathbf{p}}{(2\pi)^3}\int\frac{{\rm d}^3\mathbf{q}}{(2\pi)^3 2q^0}\sum_{i\,=\,1,2} 2\pi\delta(p_i^0-k^0-q^0) \epsilon_i \frac{|{\cal M}_i|^2}{2k^0} \nonumber\\
  && {} \times [1+f_\phi(q^0)-f_\ell(k^0)]\delta f_{pi}(0)\,\frac{1-e^{-\Gamma_i t}}{\Gamma_i}\;,
\end{eqnarray}
where the integral is over all momentum modes of the neutrino and Higgs fields (with $q^0=|{\bf q}|$), $\delta f_{pi}(0)\equiv \delta f_{i}(\mathbf{p},0)\equiv f_{i}(\mathbf{p},0)-f_{i}^{\rm eq}(\mathbf{p})$
is the initial deviation of the neutrino distribution from equilibrium, $f_\phi(q^0)=1/(e^{q^0/T}-1)$, $f_\ell(k^0)=1/(e^{k^0/T}+1)$,
and $|{\cal M}_i|^2=4k_\mu p_i^\mu (\lambda^\dag\lambda)_{ii}$ is the tree-level matrix element for the decay $N_i\to \ell\phi$.
 Furthermore, $p_i^0=\sqrt{\mathbf{p}^2+M_i^2}$ and $\Gamma_i=M_i(\lambda^\dag\lambda)_{ii}/(8\pi)$
are the energy and width of the Majorana neutrino $N_i$,
and the lepton momentum is ${\bf k}=\bf{p}-\bf{q}$ with energy $k^0=|\bf{k}|$. Finally, the asymmetry is proportional to the CP asymmetry
\begin{equation}
  \epsilon_i\ =\ \frac{1}{(\lambda^\dag\lambda)_{ii}}\, \frac{\mbox{Im}[(\lambda^\dag\lambda)_{ji}^2]}{8\pi}\frac{M_iM_j(M_j^2-M_i^2)}{(M_j^2-M_i^2)^2+A^2}\,,
\end{equation}
with a regulator $A$ that describes the saturation of the resonant enhancement. For our numerical examples, we will use
$A=M_i\Gamma_i-M_j\Gamma_j$~\cite{Anisimov:2005hr}. We stress that this is chosen for illustrative purposes only, and a
correct treatment for very small mass splitting requires to go beyond the Boltzmann approach. 
For $T\ll M_i$, the quantum-statistical corrections $1+f_\phi(q^0)-f_\ell(k^0)\to 1$ can be neglected
and the final asymmetry $Y_L^\infty=n_L(t\to\infty)/s$, normalized to the entropy density $s$, is given by 
\begin{equation}
  Y_L^\infty \ =\  \epsilon_1 \delta Y_{N_1}(0)\:+\: \epsilon_2 \delta Y_{N_2}(0)\;,
\end{equation}
where $\delta Y_{N_i}(0)=\frac{1}{s}\int\frac{{\rm d}^3\mathbf{p}}{(2\pi)^3}\,\delta f_{pi}(0)$ is the initial deviation from equilibrium.

\paragraph{Two-time Kadanoff-Baym approach:} This approach is valid for all values of the mass splitting $\Delta M=M_2-M_1$. In general,
solutions can only be obtained numerically, but analytic results valid in specific limits can be found in Ref.~\cite{Garny:2011hg}.
In particular, it is possible to obtain a semi-analytic result based on a Breit-Wigner approximation for the retarded and advanced propagators.
In this case, the total lepton asymmetry is given by
\begin{equation}\label{eq:Boltzmann}
  n_L(t) \ =\  \int\frac{{\rm d}^3\mathbf{p}}{(2\pi)^3}\int\frac{{\rm d}^3\mathbf{q}}{(2\pi)^3 2q^0}\sum_{i,j\,=\,1,2}\sum_{\epsilon\,=\,\pm}F_{ji}^\epsilon L_{ij}^\epsilon(t)\;,
\end{equation}
where the time-evolution is given by 
\begin{eqnarray}
  &&L_{ij}^\epsilon(t) \ =\ \frac{i\epsilon}{\omega_{pi}-\omega_{pj}\:+\:i\epsilon(\Gamma_{pi}+\Gamma_{pj})/2}\left[1-e^{i\epsilon(\omega_{pi}-\omega_{pj})t-(\Gamma_{pi}+\Gamma_{pj})t/2}\right] \\
 && {} \qquad \times \left(\frac{\Gamma_{\ell\phi}/2}{(\omega_{pj}-k^0-q^0+\frac{i\epsilon}{2} \Gamma_{pj})^2+\frac14\Gamma_{\ell\phi}^2}
 \:+\: \frac{\Gamma_{\ell\phi}/2}{(\omega_{pi}-k^0-q^0-\frac{i\epsilon}{2} \Gamma_{pi})^2+\frac14\Gamma_{\ell\phi}^2} \right) \nonumber\,,
\end{eqnarray}
in which $\Gamma_{\ell\phi}=\Gamma_\ell+\Gamma_\phi$ is the sum of lepton and Higgs widths, and $\Gamma_{pi}$ and $\omega_{pi}$
are the width and energy of the Majorana neutrinos, related to the imaginary and real parts of the poles of the retarded and
advanced propagators (see Ref.~\cite{Garny:2011hg} for explicit expressions). The coefficients $F_{ji}^\epsilon$ encode the initial
conditions for the Majorana neutrino two-point function $\delta S=S-S^{\mathrm{eq}}$, and its CP-conjugate $\delta\bar S$, at $t=t'=0$:
\begin{equation}
  F_{ji}^\epsilon\ =\ \mbox{tr}\left[\delta S_{Nkl}^F(0,0,\mathbf{p})\gamma^{lk\epsilon}_{ji}\:-\:\delta \bar S_{Nkl}^F(0,0,\mathbf{p})\bar \gamma^{lk\epsilon}_{ji}\right]\,,
\end{equation}
where $\gamma^{lk\epsilon}_{ji}$ and $\bar \gamma^{lk\epsilon}_{ji}$ are related to the Breit-Wigner solution of the
advanced and retarded propagators \cite{Garny:2011hg}. The flavor off-diagonal contributions ($i\not= j$) exhibit oscillations that are
crucial for the saturation of the asymmetry for very small mass splitting.

In the narrow-width limit,
\begin{equation}
  \frac{\Gamma_{\ell\phi}}{(\omega_{pi}-k^0-q^0\pm\frac{i\epsilon}{2} \Gamma_{pi})^2+\frac14\Gamma_{\ell\phi}^2}\ \to\ 2\pi\delta(\omega_{pi}-k^0-q^0)\,,
\end{equation}
one recovers the energy-conserving delta function. We consider this limit for the numerical examples that follow (see Ref.~\cite{Anisimov:2010dk} for
a discussion of finite-width effects).

For large mass splitting $\omega_{pi}-\omega_{pj}\gg \Gamma_{pi}+\Gamma_{pj}$, the flavor off-diagonal contributions $L_{ij}^\epsilon(t)$ with $i\not= j$
are suppressed relative to the diagonal ones. Furthermore, for $\Delta M/\bar M \gg  \mbox{Re}(\lambda^\dag\lambda)_{ij}/(8\pi)$ and $|{\bf p}|\lesssim M_i$, 
the energy $\omega_{pi}^2\to M_i^2+{\bf p}^2$ and 
width $\Gamma_{pi}\to \Gamma_i$ coincide with the expressions appearing in the Boltzmann results, such that $L_{ii}^\pm(t)$ has the same time-dependence
as in Eq.~(\ref{eq:Boltzmann}) and oscillations are suppressed. In Ref.~\cite{Garny:2011hg}, it has been shown that the Kadanoff-Baym approach recovers the Boltzmann result for large mass splitting and when choosing an initial condition
\begin{equation}
  \delta S_{ij}^F(0,0,\mathbf{p}) \ =\ -\,\delta_{ij}\delta f_{pi}(0)\,\frac{M_i-{\bf p}\cdot\bm{\gamma}}{\omega_{pi}}\;,
\end{equation}
where $\gamma=(\gamma^1,\gamma^2,\gamma^3)$ are the spatial Dirac gamma matrices.

In Fig.~\ref{fig:asymSplitting}, we show a comparison of the time-evolution of the lepton asymmetry for three different mass splittings,
assuming vacuum initial conditions $\delta f_{pi}(0)=-\,f_{\rm FD}(\omega_{pi})$, where $f_{\rm FD}$ is the Fermi-Dirac distribution function.
For small mass splitting $\Delta M/\bar{M} \ll \mbox{Re}(\lambda^\dag\lambda)_{ij}/(8\pi)$ (left panel), the oscillations are over-damped and the final asymmetry is suppressed compared to the Boltzmann
result. For larger mass splitting, the oscillations are visible (middle panel), and for a very large mass splitting the Kadanoff-Baym
and Boltzmann results agree very well (right panel).

\paragraph{Wigner-space approach:} The results obtained in the Wigner-space approach for the same initial conditions, and
using the simplifications applicable for $T\ll M_1$ described in Ref.~\cite{Garbrecht:2014aga}, are shown by the blue dotted lines
in Fig.~\ref{fig:asymSplitting}. The Wigner-space formulation discussed above is applicable for $\Delta M \ll \bar M$,
and we find very good agreement with the two-time Kadanoff-Baym results in this regime, when taking the narrow-width limit (left and
middle panels in Fig.~\ref{fig:asymSplitting}).
 We note that, for the chosen parameters,
the off-diagonal Yukawa coupling squared $(\lambda^\dag\lambda)_{12}$ is comparable to the diagonal couplings.
We checked that the agreement is also independent of the choice of the initial values $f_{pi}(0)$.
As expected, the Wigner-space formulation discussed above breaks down for $M_2 \gg M_1$, i.e.~for a strongly
hierarchical mass spectrum (right panel in Fig.~\ref{fig:asymSplitting}).

The final value of the asymmetry is shown in Fig.~\ref{fig:asymSplitting} as a function
of the mass splitting. One observes that the full two-time Kadanoff-Baym results can be
well approximated by either the Wigner-space result, or the Boltzmann results, depending on the
mass splitting. The Boltzmann treatment can be used for $\Delta M/\bar{M} \gg \mbox{Re}(\lambda^\dag\lambda)_{ij}/(8\pi)$, 
while the Wigner-space formulation is applicable for $\Delta M/\bar{M} \ll 1$.

It is possible to obtain an analytic result for the final value of
the asymmetry within the simplified setup studied here and for $T\ll M_1$. It is based on an analytic solution of
the evolution equation (\ref{eq:dn0hij}) for the matrix of number densities $\delta n_{hij}(t)$ (in flavor space) that describes the deviation of the quasi-degenerate Majorana
neutrinos from equilibrium for $|\mathbf{p}|\ll M_i$,
 with $a\to 1$, $f_{\rm eq}'\to 0$ and $\Gamma_{hij}\to g_w\bar M\mbox{Re}\,(\lambda^\dag\lambda)_{ij}/(32\pi)$.
Equation~\eqref{eq:dn0hij} then turns into an ordinary linear differential equation for the four components of $\delta n_{0hij}$
with constant coefficients, being the same for both helicities $h=\pm$.
The four independent solutions $\propto e^{i\Omega_{\epsilon\epsilon'} t}$ have eigen-frequencies given by 
$\Omega_{\epsilon\epsilon'}=i(\Gamma_{h11}+\Gamma_{h22}+\epsilon W_{\epsilon'}/2)$ with
\begin{eqnarray}
 W_\pm^2 \ &=&\ 2(\Gamma_{h11}-\Gamma_{h22})^2\:+\:8\Gamma_{h12}^2-2\Delta M^2\pm 2S\,, \nonumber\\
 S^2 \ &=&\ \big[(\Gamma_{h11}-\Gamma_{h22})^2-4\Gamma_{h12}^2+\Delta M^2\big]^2\:+\: 16(\Gamma_{h11}-\Gamma_{h22})^2\Gamma_{h12}^2\;.
\end{eqnarray}
For large mass splitting, they approach the values $i\Gamma_{h11}$, $i\Gamma_{h22}$ and $\pm \Delta M$, corresponding to two decaying
and two oscillating modes. When the mass splitting becomes smaller, the oscillating modes acquire a significant imaginary part
and, at some point, the real part vanishes (overdamped regime). We checked that the time-dependence agrees with the one found
in the Kadanoff-Baym approach as long as the neutrinos are weakly coupled, i.e.~$\Gamma_{hij}/\bar M\ll 1$.
Inserting this solution into the source term \eref{src:nrapp} and integrating over time, one obtains the final asymmetry.
Allowing also for flavor off-diagonal initial conditions, we find for the final yield
\begin{eqnarray}
Y_L^\infty\ =\ \epsilon^{{\rm eff}}_{11} \delta Y_{11}(0) \:+\: \epsilon^{{\rm eff}}_{22} \delta Y_{22}(0) \:+\: \epsilon^{{\rm eff}}_{12} (\delta Y_{12}(0)+\delta Y_{21}(0))\,,
\end{eqnarray}
where $\delta Y_{ij}(0)=\frac{1}{s}\int\frac{{\rm d}^3\mathbf{p}}{(2\pi)^3}\,\delta f_{0h,ij}(\mathbf{p},t=0)$ is the initial deviation of the matrix of densities from equilibrium
(assumed to be equal for both helicities $h=\pm$, as appropriate for $T\ll M_i$ \cite{Garbrecht:2014aga}) and
\begin{equation}\label{eq:cpeffweak}
  \epsilon^{{\rm eff}}_{ij} \ \equiv\  \frac{1}{\mbox{Re}(\lambda^\dag\lambda)_{ij}}\, \frac{\mbox{Im}[(\lambda^\dag\lambda)_{21}^2]}{8\pi}\frac{\bar M^2(M_2^2-M_1^2)}{(M_2^2-M_1^2)^2+A_{\rm eff}^2}\,,
\end{equation}
with 
\begin{equation}
  A_{\rm eff}\ \equiv\ \frac{\bar M^2}{8\pi}\left((\lambda^\dag\lambda)_{11}+(\lambda^\dag\lambda)_{22}\right)\left(1-\frac{[\mbox{Re}(\lambda^\dag\lambda)_{12}]^2}{(\lambda^\dag\lambda)_{11}(\lambda^\dag\lambda)_{22}}\right)^{1/2}\;.
\end{equation}
One observes that, for $\delta Y_{12}(0)=\delta Y_{21}(0)=0$ and $\delta Y_{ii}(0)=\delta Y_{N_i}(0)$, this result coincides with the Boltzmann result 
in the degenerate limit $\delta M\ll \bar M$ on replacing the regulator via $A \to A_{\rm eff}$. Furthermore, in the narrow-width limit, it coincides 
with the approximate analytic result found in the two-time Kadanoff-Baym approach in Ref.~\cite{Garny:2011hg} within its region of validity
(namely $\mbox{Re}(\lambda^\dag\lambda)_{12}\ll (\lambda^\dag\lambda)_{ii}$). The numerical results discussed above show that this agreement extends
to the case of large off-diagonal Yukawa couplings when using the full two-time Kadanoff-Baym result. 
In addition, this form of the regulator has also been found within the flavor-covariant formalism 
developed in Ref.~\cite{Dev:2014laa}, cf.~Eq.~(5.21) therein.

For given Yukawa couplings, the maximal enhancement occurs for $M_2^2-M_1^2 = A_{\rm eff}$ and is given by
\begin{equation}
  \epsilon^{{\rm eff}}_{ij}\Big|_{\rm max}\  =\  \frac{1}{\mbox{Re}(\lambda^\dag\lambda)_{ij}}\, \frac{\mbox{Im}[(\lambda^\dag\lambda)_{21}^2]}{2\left((\lambda^\dag\lambda)_{11}+(\lambda^\dag\lambda)_{22}\right)}\left(1-\frac{[\mbox{Re}(\lambda^\dag\lambda)_{12}]^2}{(\lambda^\dag\lambda)_{11}(\lambda^\dag\lambda)_{22}}\right)^{-1/2}\,.
\end{equation}
Note that the Cauchy-Schwarz inequality $|(\lambda^\dag\lambda)_{12}|^2\leq (\lambda^\dag\lambda)_{11}(\lambda^\dag\lambda)_{22}$ implies that the resonant enhancement
is well-behaved for all choices of Yukawa couplings. In particular, defining $x \equiv \mbox{Re}(\lambda^\dag\lambda)_{12}/((\lambda^\dag\lambda)_{11}(\lambda^\dag\lambda)_{22})^{1/2}$, the Cauchy-Schwarz inequality implies $|x|\leq 1$ and $(\mbox{Im}(\lambda^\dag\lambda)_{21})^2\leq (\lambda^\dag\lambda)_{11}(\lambda^\dag\lambda)_{22}(1-x^2)$.
We therefore set $y \equiv \mbox{Im}(\lambda^\dag\lambda)_{21}/((\lambda^\dag\lambda)_{11}(\lambda^\dag\lambda)_{22}(1-x^2))^{1/2}$, implying also $|y|\leq 1$.
In terms of these parameters
\begin{subequations}
\begin{align}
  \epsilon^{{\rm eff}}_{ii}\Big|_{\rm max}\ &=\ xy\,\frac{(\lambda^\dag\lambda)_{ii}}{(\lambda^\dag\lambda)_{11}+(\lambda^\dag\lambda)_{22}}\,,\\
  \epsilon^{{\rm eff}}_{12}\Big|_{\rm max}\ &=\ y\,\frac{((\lambda^\dag\lambda)_{11}(\lambda^\dag\lambda)_{22})^{1/2}}{(\lambda^\dag\lambda)_{11}+(\lambda^\dag\lambda)_{22}}\;.
  \end{align}
\end{subequations}
The absolute values of the flavor-diagonal and off-diagonal contributions are therefore bounded to be below $1$ and $1/2$, respectively.
Note also that $\epsilon^{{\rm eff}}_{\rm av}\equiv(\epsilon^{{\rm eff}}_{11}+\epsilon^{{\rm eff}}_{22})/2\leq 1/2$.

Finally, we caution that the actual time-evolution of the asymmetry is very different from the Boltzmann result for $M_2^2-M_1^2 \lesssim \mbox{\ few\ }\times A_{\rm eff}$.
In particular, flavor off-diagonal correlations are built up even when they are vanishing initially and are crucial to capture the saturation of the enhancement.
Therefore, it is necessary to go beyond the standard Boltzmann treatment, for instance by solving the two-time Kadanoff-Baym or the Wigner-space evolution equations, in order to obtain an accurate result for the lepton asymmetry
in situations that differ from the simplified setup considered here. The two-momentum approach has been compared to the Kadanoff-Baym and Wigner descriptions for
a scalar model in Ref.~\cite{Kartavtsev:2015vto}, finding agreement if the initial conditions are properly related, and up to terms suppressed by $\Delta M/\bar M$ (see Sec.~\ref{sec:scalar}).
This approach allows one to separate the contribution to the asymmetry into contributions ascribed to oscillations and to mixing, as we will describe in the next section. A similar separation is possible
within the Wigner approach, when considering the contributions to the asymmetry from the various eigenmodes discussed above separately. For further discussions of the sources of CP asymmetry, see the companion Chapter~\cite{leptogenesis:A01}.

\begin{figure}[!t]
\centering
\includegraphics[width=0.3\textwidth]{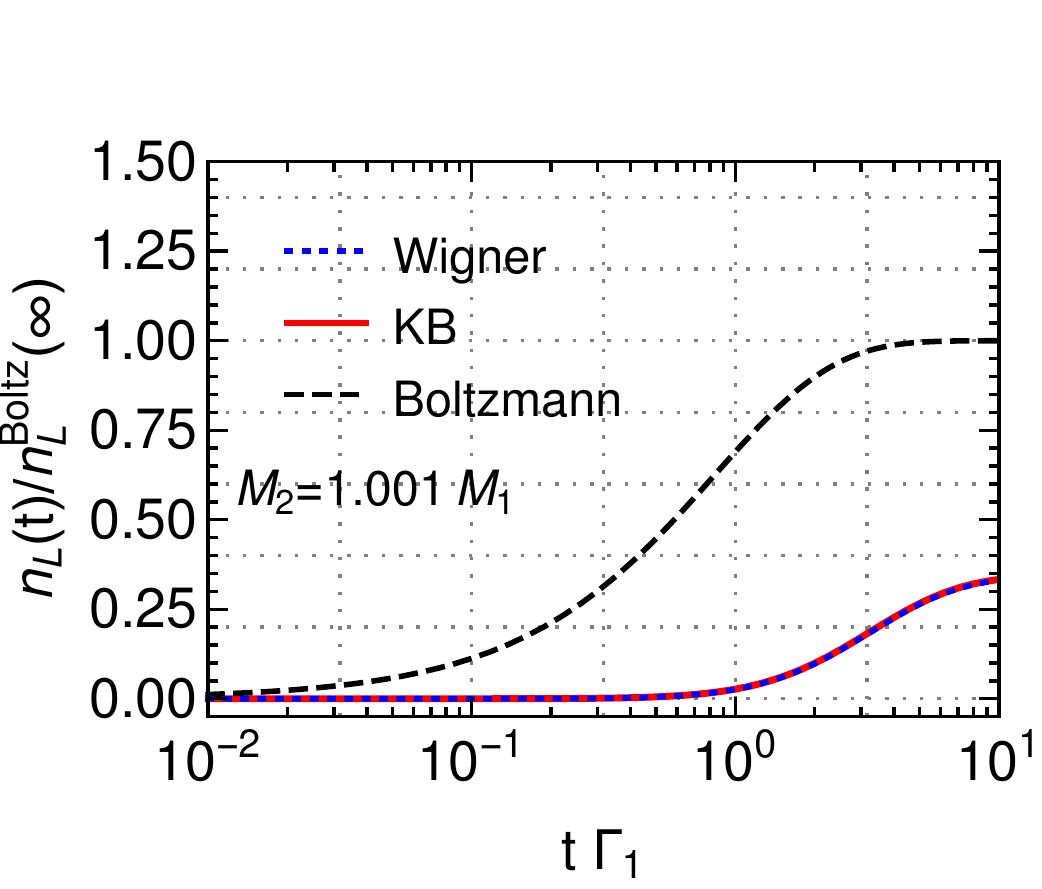}
\includegraphics[width=0.3\textwidth]{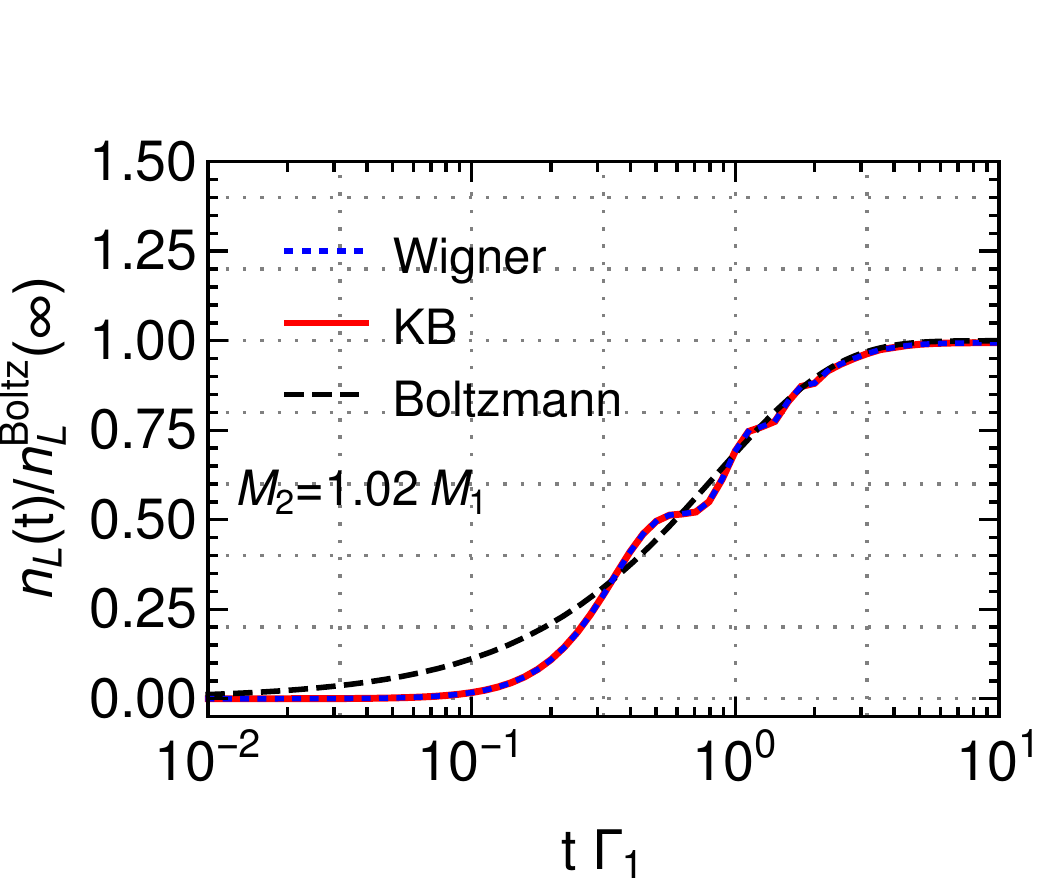}
\includegraphics[width=0.3\textwidth]{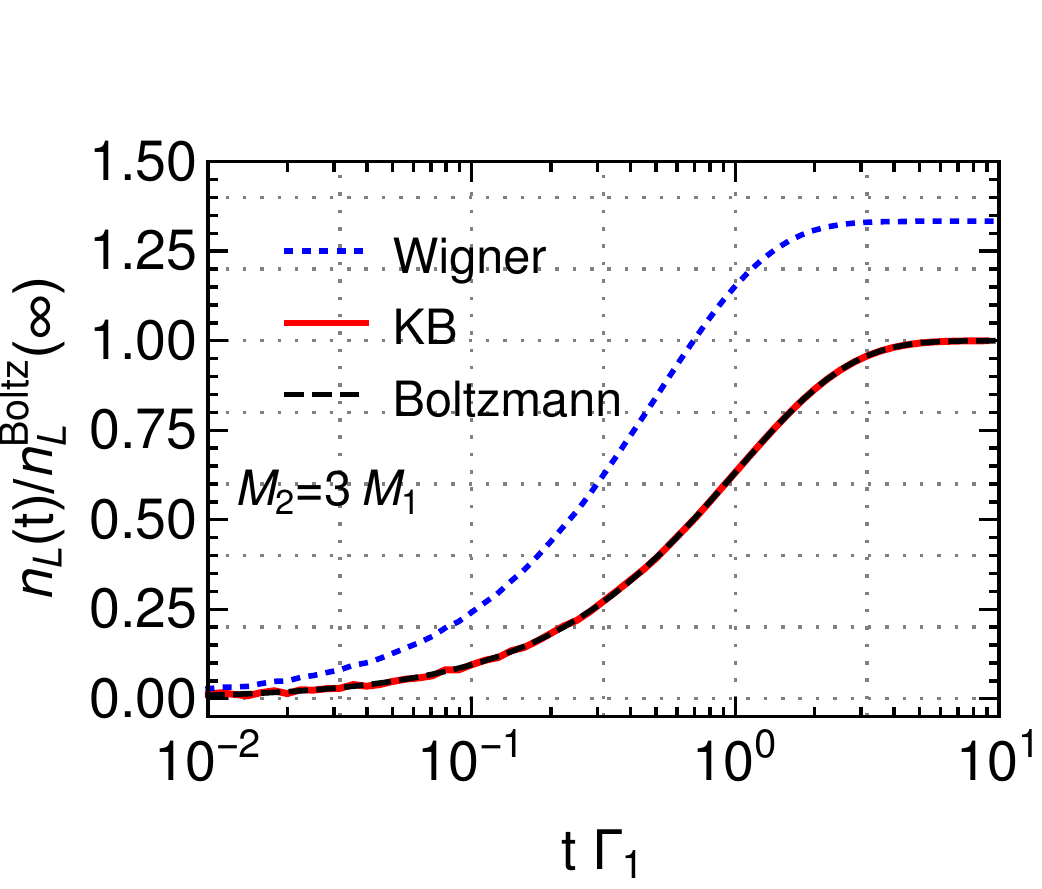}
\caption{\label{fig:asym}
Time-evolution of the lepton asymmetry for three different mass splittings $M_2=(0.001,1.02,3)M_1$, respectively. Each figure shows the
result obtained in the Wigner formulation (blue dotted), the two-time Kadanoff-Baym approach (red) and, for comparison, the conventional
Boltzmann description (black dashed). Parameters are $\lambda^\dag\lambda=0.05\, ((1, e^{i\pi/8}), (e^{-i\pi/8}, 1.5))$, $T=M_1/10$, and the time axis
is in units of $\Gamma_1=M_1(\lambda^\dag\lambda)_{11}/(8\pi)$.}
\end{figure}

\begin{figure}[!t]
\begin{center}
\includegraphics[width=0.75\textwidth]{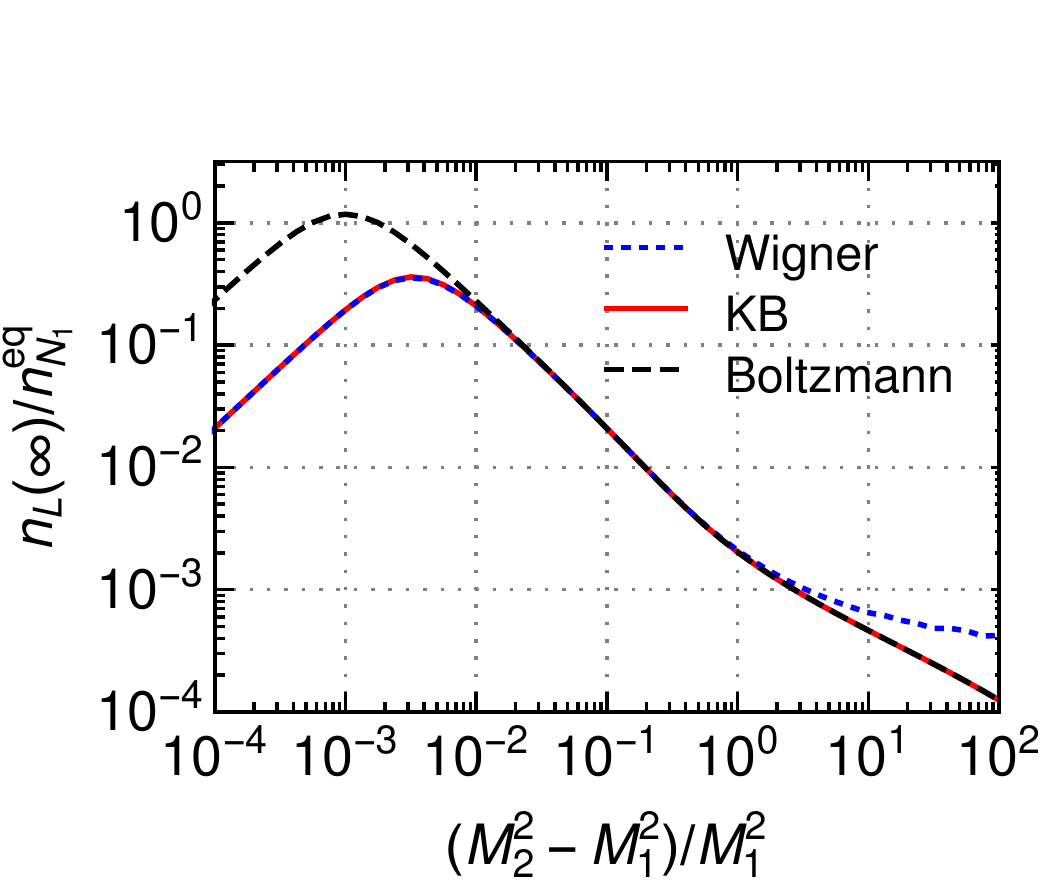}
\end{center}
\caption{\label{fig:asymSplitting}
Final lepton asymmetry produced in the relaxation of $N_{1,2}$ from initial vacuum abundance into thermal equilibrium, depending
on the mass splitting of the $\overline{\rm MS}$ masses $M_{1,2}$. The color code and parameters are chosen as in
Fig.~\ref{fig:asym}}
\end{figure}

\section{Interaction picture versus two-time formulation}\label{sec:scalar}

As mentioned above, a full field-theoretic implementation of the two-momentum formulation based on the interaction picture
has been studied in Ref.~\cite{Dev:2014wsa} in the context of a scalar-field model. Below we review the main results, and comment on
a comparison to the two-time formulation following Ref.~\cite{Kartavtsev:2015vto}.

We can construct the following and particularly useful toy model of RL. It comprises two real scalar fields $N_{k}$ ($k=1,2$), which model the two lightest heavy neutrinos, and one complex scalar field $\ell$, which mimics a single generation of charged leptons and whose associated $U(1)$ symmetry models lepton number $L$. The Lagrangian has the form
\begin{equation}\label{eq:Ltoy}
-\mathcal{L}\ = \ \frac{1}{2}\lambda_i^* \ell^{\dag} \ell^{\dag} N_i\:+\:\frac{1}{4}\,N_i(M^2)_{ij}N_j\:+\:\mathrm{h.c.}\;,
\end{equation}
where $\lambda_i$ model the tree-level Yukawa couplings\footnote{We draw attention to the complex conjugation of the would-be Yukawa couplings in Eq.\,\eqref{eq:Ltoy} relative to the realistic Yukawa couplings in Eq.\,\eqref{eq:L}}. The would-be lepton number is broken by the $\ell^{\dag}\ell^{\dag} N$ term (and its hermitian conjugate), and C is violated (along with CP) as long as the would-be heavy neutrinos are non-degenerate and $\mathrm{arg}\,\lambda_1\neq \mathrm{arg}\,\lambda_2$. Hence, all three Sakharov conditions \cite{Sakharov:1967dj} are satisfied when we also provide either non-equilibrium initial conditions or place the system in an expanding Universe. This toy model has been used to analyze a number of aspects of leptogenesis~\cite{Garny:2009rv,Garny:2009qn,Garny:2010nj}, including the impact of effective thermal masses~\cite{Hohenegger:2014cpa}. A similar scalar toy model, in which the term $\lambda_i^*\ell^{\dag}\ell^{\dag}N_i/2$ is replaced by $\lambda_i\ell^{\dag}\phi N_i$ (where the real scalar field $\phi$ plays the role of the SM Higgs), has also been used to make comparisons between semi-classical and field-theoretic descriptions of RL~\cite{Dev:2014wsa,Dev:2015dka}.

Working in the mass eigenbasis and for a Gaussian and spatially-homogeneous ensemble for the heavy neutrinos, the two-momentum representation of their tree-level statistical propagator takes the form~\cite{Dev:2014wsa,Kartavtsev:2015vto}
\begin{align}
\label{eq:wouldbeNprop}
&\widetilde{S}^{F,0}_{Nij}(p,p',t)\ =\ 2\pi|p_0|^{1/2}\delta(p^2-M_i^2)e^{i(p_0-p_0')t}(2\pi)^3\delta^{(3)}(\mathbf{p}-\mathbf{p}')\nonumber\\&\qquad\times\:\Big\{\theta(+p_0,+p_0')\Big[\delta_{ij}+2f_{ij}(t,\mathbf{p})\Big]+\theta(-p_0,-p_0')\Big[\delta_{ij}+2f_{ij}^*(t,\mathbf{p})\Big]\Big\}\nonumber\\&\qquad \times\:2\pi|p_0'|^{1/2}\delta(p'^2-M_j^2)\;.
\end{align}
We use a tilde to identify that this is a scalar propagator and to avoid confusion with the spinor propagators that appear elsewhere in this review. The phase $e^{i(p_0-p_0')t}$ accounts for the free evolution of the interaction-picture operators, and $f_{ij}(t,\mathbf{p})$ are the elements of the matrix of would-be heavy-neutrino distribution functions. We have also defined $\theta(x,y)\equiv \theta(x)\theta(y)$. We immediately see the advantage of the two-momentum formulation: the spectral structure of the off-diagonal elements ($i\neq j$) is such that on-shell four-momentum $p'^2=M_j^2$ flows in and, following an interaction with the statistical ensemble, on-shell four-momentum $p^2=M_i^2$ flows out. This tree-level spectral structure resembles the composite structure for the Wightman functions obtained by means of the coherent quasi-particle approximation (cQPA) in Refs.~\cite{Herranen:2010mh,Herranen:2011zg,Fidler:2011yq}.

It is illustrative to first take the Wigner transform of Eq.~\eqref{eq:wouldbeNprop} by defining the relative and central momenta $Q^{\mu}=p^{\mu}-p^{\mu\prime}$ and $q^{\mu}=(p^{\mu}+p^{\mu\prime})/2$. Specifically, we find (in the equal-time limit $X_0=t$)
\begin{align}
\label{eq:treeSFtilde}
\widetilde{S}^{F,0}_{Nij}(q,t)\ &=\ \int\frac{{\rm d}^4Q}{(2\pi)^4}\;e^{-i(Q_0 t-\mathbf{Q}\cdot\mathbf{X})}\;\widetilde{S}^{F,0}_{Nij}(q+Q/2,q-Q/2,t)\nonumber\\&
=\ \pi\,\frac{E_{ij}}{(E_iE_j)^{1/2}}\, \delta(q^2_0-E^2_{ij}(\mathbf{q}))\nonumber\\&\qquad\times\:\Big\{\theta(+q_0)\big(\delta_{ij}+2f_{ij}(t,\mathbf{q})\big)\:+\:\theta(-q_0)\big(\delta_{kl}+2f_{ij}^*(t,\mathbf{q})\big)\Big\}\;,
\end{align}
where $E_{ij}\equiv (E_i+E_j)/2$ and $E_{i,j}\equiv E_{i,j}(\mathbf{q})=\big(\mathbf{q}^2+M_{i,j}^2\big)^{1/2}$. We see immediately that the spectral structure comprises three shells (see also Ref.~\cite{Fidler:2011yq}): two associated with the on-shell energies $E_1$ and $E_2$, and one associated with the $\bar{E}=(E_1+E_2)/2$. Much of the delicacy of the treatment of RL is related to how this three-shell structure is modified by the resummation of self-energy corrections, and it is to this aspect that we now turn our attention.

In a Markovian approximation, and assuming that the self-energies and retarded and advanced propagators are translationally invariant in the thermal bath, one-loop self-energy corrections can be resummed in closed form in the two-momentum representation~\cite{Dev:2014wsa,Kartavtsev:2015vto}. Doing so, one obtains the following result:
\begin{align}
\label{eq:wouldbeNpropresum}
\widetilde{S}^F_{Nij}(p,p',t)\ &=\ 	F^R_{ik}(p)\widetilde{S}^{F,0}_{Nkl}(p,p',t)F^A_{lj}(p')\nonumber\\
&\qquad -\:\widetilde{S}^R_{Nik}(p)\widetilde{\Sigma}^F_{Nkl}(p)(2\pi)^4\delta^{(4)}(p-p')\widetilde{S}^A_{Nlj}(p')\;,
\end{align}
where $\widetilde{S}^{R(A)}_{Nij}(p)$ is the resummed retarded (advanced) heavy-neutrino propagator and
\begin{subequations}
\begin{align}
F^{R}_{ij}(p)\ =\ \sum_{n\,=\,0}^{\infty}\Big[(-\widetilde{S}_N^{R,0}\cdot\widetilde{\Sigma}_N^R)^n\Big]_{ij}\ =\ -\,\widetilde{S}^{R}_{Nik}[\widetilde{S}_N^{R,0}]^{-1}_{kj}\;,\\
F^{A}_{ij}(p)\ =\ \sum_{n\,=\,0}^{\infty}\Big[(-\widetilde{\Sigma}_N^A\cdot\widetilde{S}_N^{A,0})^n\Big]_{ij}\ =\ -\,[\widetilde{S}_N^{A,0}]^{-1}_{ik}\widetilde{S}^{A}_{Nkj}\;.
\end{align}
\end{subequations}
Whilst the first term on the rhs of Eq.~\eqref{eq:wouldbeNpropresum} contributes to the source term for the asymmetry, the right-most term in Eq.~\eqref{eq:wouldbeNpropresum} describes equilibrium $\Delta L=0$ and $\Delta L=2$ scatterings and is relevant only to the washout terms.

Proceeding to analyze only the source term in the weak washout regime and in a Minkowski space-time background, connection of the above resummation with the semi-classical treatment involving effective, resummed Yukawa couplings (see e.g.~Refs.~\cite{Pilaftsis:2003gt,Dev:2014laa}) can be made by virtue of the following equivalence in the heavy-neutrino mass eigenbasis:
\begin{equation}
\label{eq:effYukawaidentity}
\lambda_iF^R_{ij}(p)\ \sim \ \widehat{\lambda}_i\;,
\end{equation}
as identified in Ref.~\cite{Dev:2014wsa}, where $\widehat{\lambda}_i$ is the resummed Yukawa coupling. When evaluated on the $i$-th mass shell, the resummed Yukawa couplings and their $C$-conjugates are given by~\cite{Garny:2009qn,Kartavtsev:2015vto}
\begin{equation}
\widehat{\lambda}_i^{(c)}\ =\ \lambda_i\bigg[1-(+)i\frac{\lambda_i\lambda_{\slashed{i}}^*}{32\pi}\bigg(1+\frac{\lambda_i^*\lambda_{\slashed{i}}}{\lambda_i\lambda_{\slashed{i}}^*}\bigg)\frac{1}{\Delta M_{i\slashed{i}}^2+(-)iM_{\slashed{i}}\Gamma_{\slashed{i}}}\bigg]\;,
\end{equation}
where $\Gamma_{i}=|\lambda_i|^2/(16\pi M_i)$ is the tree-level decay width of the $i$-th heavy neutrino. Here, we use a shorthand notation where $\slashed{i}=1$ when $i=2$ and vice versa~\cite{Dev:2014wsa}.

Since the heavy neutrinos are unstable, they cannot appear as asymptotic in or out states. The resummed Yukawa couplings are therefore calculated by considering the $S$-matrix elements of (the would-be) charged-lepton scatterings and treating carefully the pole and residue structure of the intermediate heavy neutrinos. In this way, one can obtain a resummation scheme that preserves important field-theoretic properties, such as unitarity and CPT invariance. For more detailed discussions of the resummation approaches employed in semi-classical analyses, see Refs.~\cite{Pilaftsis:1998pd} and \cite{Pilaftsis:2003gt}, and references therein, as well as \sref{sec:earlyhistory}.

Substituting for Eq.~\eqref{eq:effYukawaidentity} into the non-equilibrium part of the resummed statistical propagator, Eq.~\eqref{eq:wouldbeNpropresum}, it can be shown that the time-derivative of the asymmetry can be written in the form\cite{Kartavtsev:2015vto}
\begin{align}
\label{eq:asymmetyeffective}
\frac{{\rm d} n_{L}}{{\rm d}t}\ &\sim \ 2\sum_i\int\!\frac{{\rm d}^3\mathbf{q}}{(2\pi)^3}\,\frac{M_{i}}{\bar{\omega}}\,\delta f_{ii}(t,\mathbf{q})\Gamma_{i}^{\rm med}(\bar{\omega},\mathbf{q})\epsilon^{\rm vac}_i\nonumber\\&\qquad +\:2\,\mathrm{Im}(\lambda_1\lambda_2^*)\int\!\frac{{\rm d}^3\mathbf{q}}{(2\pi)^3}\,\frac{\widetilde{\Sigma}_N^{\rho}(\bar{\omega},\mathbf{q})}{\bar{\omega}}\,\mathrm{Im}\,\delta f_{12}(t,\mathbf{q})\;.
\end{align}
The various terms appearing are as follows: $\delta f_{ij}(t,\mathbf{q})\equiv f_{ij}(t,\mathbf{q})-f^{\rm eq}_{ij}(\mathbf{q})$ are the deviations from equilibrium of the elements of the heavy-neutrino matrix of number densities; $\bar{\omega}=(\omega_1+\omega_2)/2$ is the intermediate quasi-particle mass shell (where we allow for the inclusion of local dispersive corrections to the $M_i$, viz.~thermal masses); $\Gamma_{i}^{\rm med}(\bar{\omega},\mathbf{q})=\Gamma_{i}L^{\rho}(\bar{\omega},\mathbf{q})$ is the in-medium width; $\epsilon_i^{\rm vac}$ is the well-known CP-violating parameter
\begin{equation}
\label{eq:epsilonvac}
\epsilon_i^{\rm vac}\ =\ \frac{\Gamma_{N_i\to \ell\ell}-\Gamma_{N_i\to \ell^{\dag}\ell^{\dag}}}{\Gamma_{N_i\to \ell\ell}+\Gamma_{N_i\to \ell^{\dag}\ell^{\dag}}}\ =\ \mathrm{Im}\,\bigg(\frac{\lambda_i\lambda_{\slashed{i}}^*}{\lambda_i^*\lambda_{\slashed{i}}}\bigg)\,\frac{(M_{i}^2-M_{\slashed{i}}^2)M_{\slashed{i}}\Gamma_{\slashed{i}}}{(M_{i}^2-M_{\slashed{i}}^2)^2+(M_{\slashed{i}}\Gamma_{\slashed{i}})^2}\;;
\end{equation}
and
\begin{equation}
\widetilde{\Sigma}_N^{\rho}(q_0,\mathbf{q})\ =\ \frac{1}{8\pi}\,L^{\rho}(q_0,\mathbf{q})\;.
\end{equation}
The function
\begin{equation}
L^{\rho}(q_0,\mathbf{q})\ =\ 1\:+\:\frac{2T}{|\mathbf{q}|}\,\ln\bigg[\frac{1-e^{-(q_0+|\mathbf{q}|)/2T}}{1-e^{-(q_0-|\mathbf{q}|)/2T}}\bigg]\;,
\end{equation}
as given in the $\overline{\mathrm{MS}}$ scheme, accounts for the in-medium corrections (see Ref.~\cite{Hohenegger:2014cpa} and accompanying Chapter~\cite{leptogenesis:A04}). The source term itself is obtained in complete analogy to the discussion in Sec.~\ref{sec:methodsoverview} (see also Refs.~\cite{Kartavtsev:2015vto, Hohenegger:2014cpa}). We note that the resummation approach above can be generalized to finite temperature in order to capture in-medium effects in the CP-violating parameter, and this is, in fact, important for performing real-intermediate state (RIS) subtraction in the presence of off-diagonal flavor correlations~\cite{Dev:2014laa,Dev:2014wsa}.

Alternatively, we can expand the non-equilibrium part of the Wigner transform of Eq.~\eqref{eq:wouldbeNpropresum} to first order in $\widetilde{\Sigma}_N^{R(A)}$. Considering only the positive-frequency part ($q_0>0$), we have~\cite{Kartavtsev:2015vto}
\begin{align}
\label{eq:noneqparttildeS}
\widetilde{S}^F_{Ni\slashed{i}}(t,q)\ &\supset\ 2\pi\delta(q_0-\omega_i)\,\frac{1}{2\omega_i}\,\delta f_{ii}(t,\mathbf{q})\widetilde{\Sigma}^A_{Ni\slashed{i}}(\omega_i,\mathbf{q})R_{i\slashed{i}}\nonumber\\& -\:2\pi\delta(q_0-\omega_{\slashed{i}})\,\frac{1}{2\omega_{\slashed{i}}}\,\delta f_{\slashed{i}\slashed{i}}(t,\mathbf{q})\widetilde{\Sigma}^R_{Ni\slashed{i}}(\omega_{\slashed{i}},\mathbf{q})R_{i\slashed{i}}\nonumber\\&+\:2\pi\delta(q_0-\bar{\omega})\,\frac{1}{(2\omega_i)^{1/2}(2\omega_{\slashed{i}})^{1/2}}\Big[\delta f_{i\slashed{i}}(t,\mathbf{q})\,\Delta M_{i\slashed{i}}^2\nonumber\\&\qquad-\:\delta f_{ii}(t,\mathbf{q})\widetilde{\Sigma}^A_{Ni\slashed{i}}(\omega_{\slashed{i}})\:+\:\delta f_{\slashed{i}\slashed{i}}(t,\mathbf{q})\widetilde{\Sigma}^R_{Ni\slashed{i}}(\omega_i)\Big]R_{i\slashed{i}}\;,
\end{align}
where $\Delta M_{i\slashed{i}}^2\equiv M_{i}^2-M_{\slashed{i}}^2$ is the mass splitting and
\begin{equation}
R_{i\slashed{i}}\ \equiv\ \frac{\Delta M_{i\slashed{i}}^2}{(\Delta M_{i\slashed{i}}^2)^2+(\omega_i\Gamma_{i}-\omega_{\slashed{i}}\Gamma_{\slashed{i}})^2}\;.
\end{equation}
We see that there is complete cancellation of the first, second and fourth lines of Eq.~\eqref{eq:noneqparttildeS} in the limit $\omega_i=\omega_{\slashed{i}}$. It is this cancellation that ensures that the asymmetry vanishes in the degenerate limit, as it should. Moreover, by everywhere replacing $\omega_i$ and $\omega_{\slashed{i}}$ by $\bar{\omega}$, we obtain a propagator of the form
\begin{align}
\widetilde{S}^F_{Ni\slashed{i}}(t,q)\ &\supset\ 2\pi\delta(q^2-\bar{M}^2)\theta(q_0)\delta f_{i\slashed{i}}(t,\mathbf{q})\,\Delta M_{i\slashed{i}}^2R_{i\slashed{i}}\;,
\end{align}
which resembles the average mass approximation (cf.~also Eq.~\eqref{eq:treeSFtilde}) for the quasi-particle propagators, discussed in \sref{sec:wigner}, up to the regulator factor $\Delta M_{i\slashed{i}}^2R_{i\slashed{i}}$. The latter is of order unity in the weakly resonant or overlapping regime where $\Gamma_i\ll \Delta M\ll \bar{M}$.

Making use of Eq.~\eqref{eq:noneqparttildeS}, the time-derivative of the asymmetry takes the following form to leading order in $\widetilde{\Sigma}^{R(A)}_N$~\cite{Kartavtsev:2015vto}:
\begin{align}
\label{eq:scalarasym}
\frac{{\rm d} n_L(t)}{{\rm d} t}\ &\approx\ 2\sum_i\int\frac{{\rm d}^3\mathbf{q}}{(2\pi)^3}\,\frac{M_{i}}{\omega_i}\,\delta f_{ii}(t,\mathbf{q})\Gamma^{\rm med}_{i}(\omega_i,\mathbf{q})\epsilon_i^{\rm med}(\omega_i,\mathbf{q})\nonumber\\&+\:2\,\mathrm{Im}(\lambda_1\lambda_{2}^*)\,\mathrm{Im}\int\frac{{\rm d}^3\mathbf{q}}{(2\pi)^3}\,\frac{\widetilde{\Sigma}_N^{\rho}(\bar{\omega},\mathbf{q})}{(\omega_1\omega_2)^{1/2}}\Big[\delta f_{12}(t,\mathbf{q})\,\Delta M_{12}^2\nonumber\\&\qquad-\:\delta f_{11}(t,\mathbf{q})\widetilde{\Sigma}^A_{N12}(\omega_2,\mathbf{q})\:+\:\delta f_{22}(t,\mathbf{q})\widetilde{\Sigma}^R_{N12}(\omega_1,\mathbf{q})\Big]R_{12}\;,
\end{align}
where
\begin{equation}
\label{eq:epsilonmed}
\epsilon_i^{\rm med}(\omega_i,\mathbf{q})\ =\ \mathrm{Im}\,\bigg(\frac{\lambda_i\lambda_{\slashed{i}}^*}{\lambda_i^*\lambda_{\slashed{i}}}\bigg)\,\frac{(M_{i}^2-M_{\slashed{i}}^2)M_{\slashed{i}}\Gamma_{\slashed{i}}}{(\Delta M_{i\slashed{i}}^2)^2+(\omega_i\Gamma_i-\omega_{\slashed{i}}\Gamma_{\slashed{i}})^2}\,L^{\rho}(\omega_i,\mathbf{q})
\end{equation}
is the in-medium CP-asymmetry parameter (in which we have neglected the in-medium corrections to the widths in the denominator, valid away from the strongly resonant regime, cf.~Ref.~\cite{Garny:2009qn}). We associate the terms of the first line of Eq.~\eqref{eq:scalarasym}, being proportional to the diagonal entries of $f_{ij}$, with the source of asymmetry due to \emph{mixing}. These result from the contributions of the quasi-particle mass shells in the first two lines of Eq.~\eqref{eq:noneqparttildeS}. The terms of the second line of Eq.~\eqref{eq:scalarasym}, being proportional to the off-diagonal entries of $f_{ij}$, are associated with \emph{oscillations} and result from the contribution of the intermediate shell in the third line of Eq.~\eqref{eq:noneqparttildeS}. Finally, the terms in the last line of Eq.~\eqref{eq:scalarasym}, which cancel with the mixing source in the limit  $\omega_i=\omega_{\slashed{i}}$, can be interpreted as the destructive \emph{interference} between the mixing and oscillation sources. The latter live on the intermediate mass shell but are proportional to the diagonal entries of $f_{ij}$.

In order to make comparison with the effective Yukawa approach above, we can expand all but the regulator structure in $R_{12}$ around $\Delta \omega_{i\slashed{i}}=\omega_i-\omega_{\slashed{i}}=0$. Doing so, we obtain~\cite{Kartavtsev:2015vto}
\begin{align}
\label{eq:asymmetryapprox}
\frac{{\rm d} n_L(t)}{{\rm d} t}\ &\approx\ 2\sum_i\int\!\frac{{\rm d}^3\mathbf{q}}{(2\pi)^3}\,\frac{M_{i}}{\bar{\omega}}\,\delta f_{ii}(t,\mathbf{q})\Gamma^{\rm med}_{i}(\bar{\omega},\mathbf{q})\bar{\epsilon}_i^{\rm med}(\bar{\omega},\mathbf{q})\nonumber\\&+\:2\,\mathrm{Im}(\lambda_1\lambda_{2}^*)\int\!\frac{{\rm d}^3\mathbf{q}}{(2\pi)^3}\,\frac{\widetilde{\Sigma}_N^{\rho}(\bar{\omega},\mathbf{q})}{\bar{\omega}}\,\mathrm{Im}\,\delta f_{12}(t,\mathbf{q})\,\Delta M_{12}^2\,R_{12}\;.
\end{align}
Whilst this resembles the result in Eq.~\eqref{eq:asymmetyeffective} up to the factor of $\Delta M_{12}^2\,R_{12}$, which is unity in the weakly resonant regime (as discussed earlier), the CP-violating parameter has been modified~\cite{Kartavtsev:2015vto}:
\begin{equation}
\label{eq:mixmod}
\bar{\epsilon}^{\rm med}_i\ \equiv\ \frac{\omega_{\slashed{i}}-\omega_i}{\omega_{\slashed{i}}+\omega_i}\,\epsilon_i^{\rm med}\;.
\end{equation}
As we approach the degenerate limit, this would appear to suggest an additional suppression of this source of CP asymmetry. Moreover, it results in an additional sign difference between the sources from the two flavors, such that further cancellation results in scenarios where $\delta f_{11}\sim \delta f_{22}$. Comparing this with Eq.~\eqref{eq:asymmetyeffective}, it would, at first sight, appear that this destructive interference between the contributions from the different mass shells is not captured in the resummed Yukawa approach. However, setting to unity the in-medium corrections $L^{\rho}$, one notices that the structure of the CP violating parameter differs between Eqs.~\eqref{eq:epsilonvac} and \eqref{eq:epsilonmed}. In phenomenological studies (see e.g.~Refs.~\cite{Pilaftsis:2003gt, Dev:2014laa}), the form of the regulator in Eq.~\eqref{eq:epsilonvac} ensures that the asymmetry vanishes in the degenerate limit, as it should, in spite of the absence of the additional terms that were present in the fourth line of Eq.~\eqref{eq:noneqparttildeS}. In this way, this destructive interference is, at least in part, captured by the effective Yukawa approach.  

\begin{figure}[t!]
\includegraphics[width=0.49\textwidth]{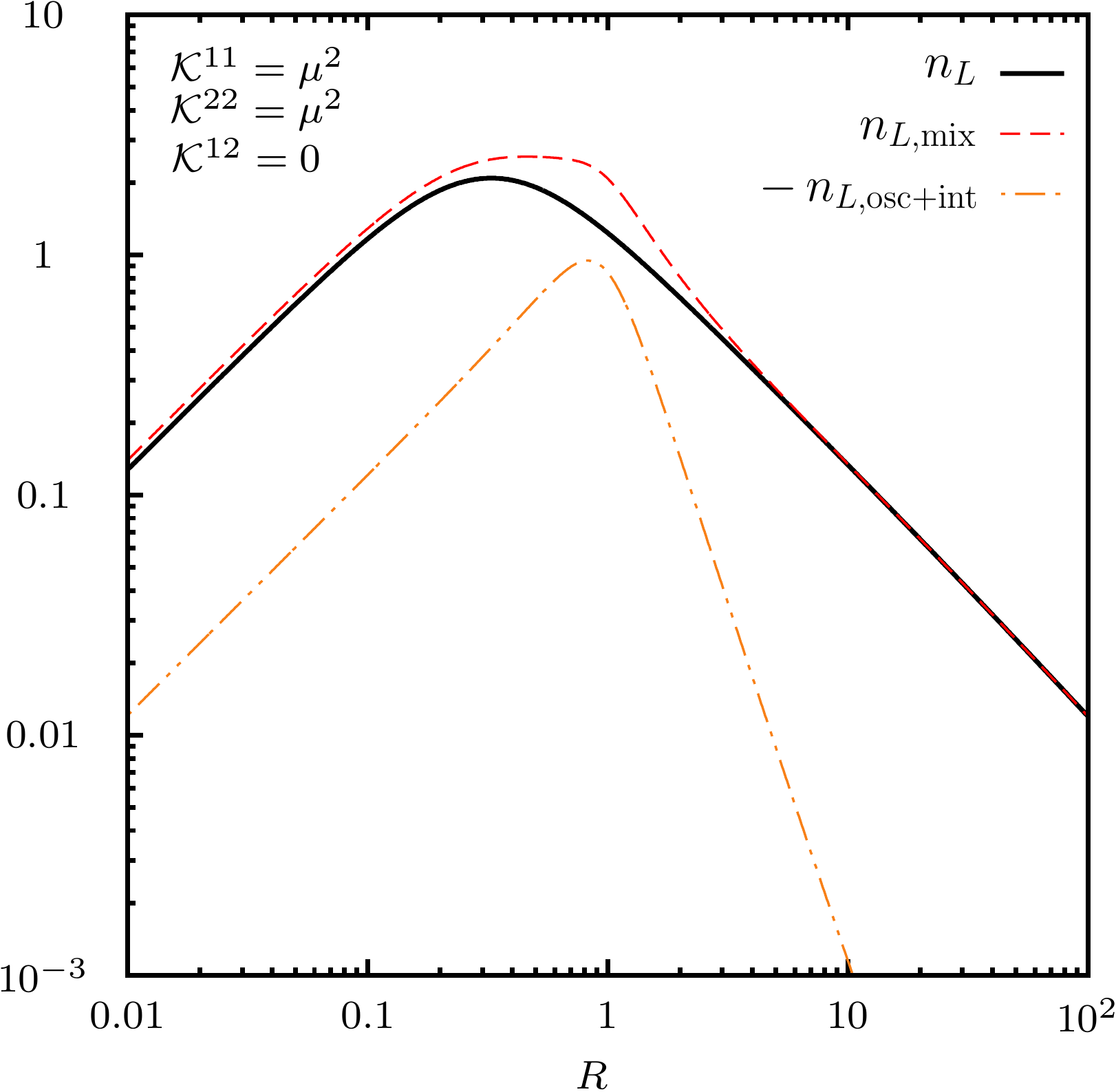} \includegraphics[width=0.49\textwidth]{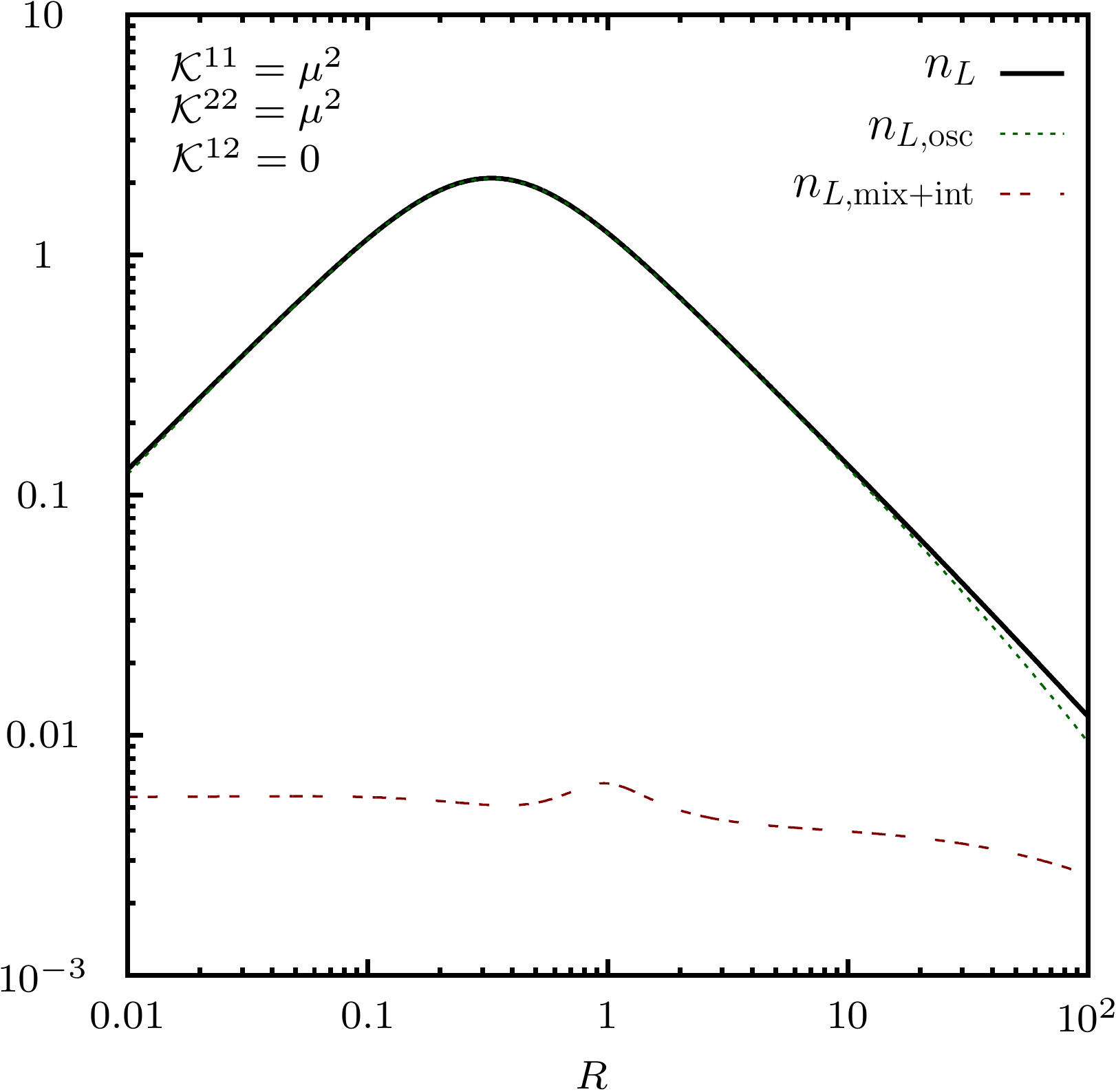}\\
\includegraphics[width=0.49\textwidth]{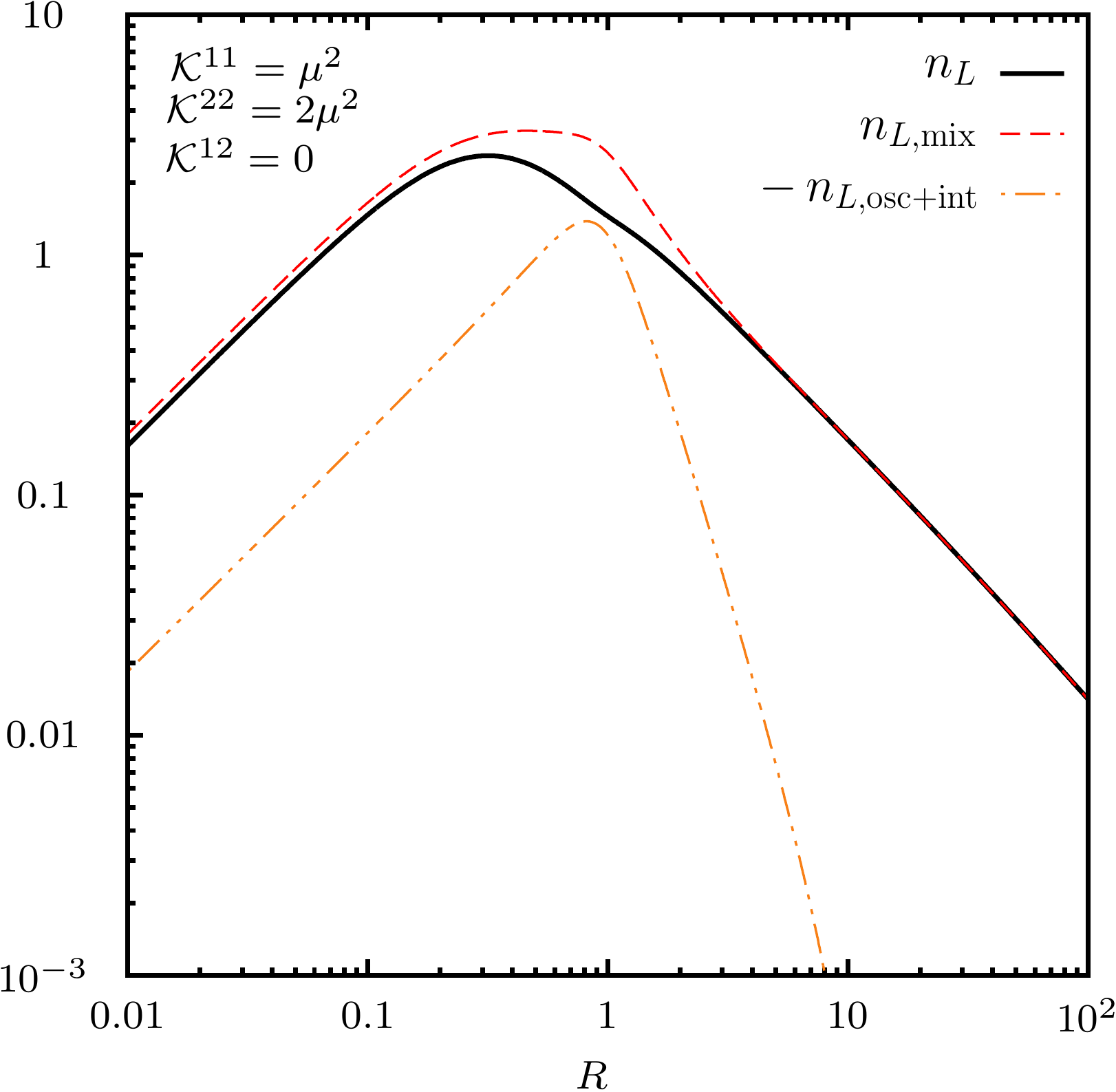} \includegraphics[width=0.49\textwidth]{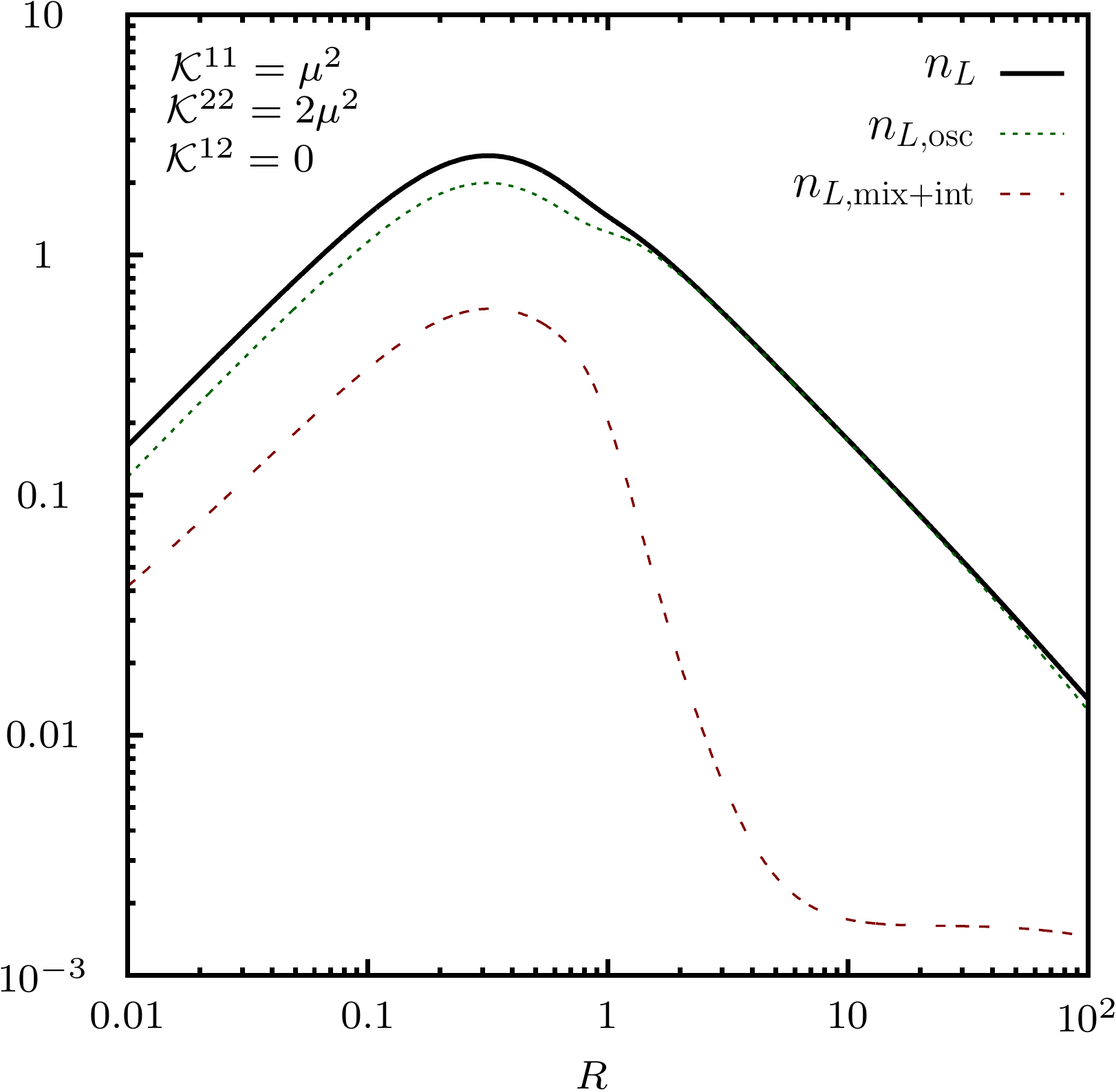}
\caption{\label{fig:mixosc}The various contributions to the total asymmetry $n_L$ interpreted in terms of  the Boltzmann benchmark $n_{L,\rm mix}$ with corrections $n_{L,\rm osc+int}$ (left panels) and the density matrix benchmark $n_{L,\rm osc}$ with corrections $n_{L,\rm mix+int}$ (right panels) as functions of the degeneracy parameter $R$ in Eq.~\eqref{eq:degenparam}
for two diagonal, C-conserving choices of the initial conditions specified in terms of the parameters $\mathcal{K}^{11}$, $\mathcal{K}^{22}$ and $\mathcal{K}^{12}$ (see Ref.~\cite{Kartavtsev:2015vto} for more details). The would-be Yukawa couplings were taken to be $h_1=0.5\,\mu\exp(-i)$ and $h_2=-\,0.8\,\mu\exp(-2i/3)$, the temperature $T=\mu$ and the mass $M_1=\mu$, where $\mu$ is the $\overline{\mathrm{MS}}$ renormalization scale. Figures adapted from Ref.~\cite{Kartavtsev:2015vto}.}
\end{figure} 

Before closing this section, we comment on the relative magnitudes of the mixing, oscillation and interference contributions as a function of the degeneracy parameter
\begin{equation}
\label{eq:degenparam}
R\ =\ \frac{M_2^2-M_1^2}{M_1\Gamma_1+M_2\Gamma_2}
\end{equation}
in the weak washout regime. Further details of the analysis, as well as the correct specification of ${\rm C}$-symmetric initial conditions, can be found in Ref.~\cite{Kartavtsev:2015vto}. The results are shown in Fig.~\ref{fig:mixosc}, and we draw the following conclusions:
\begin{itemize}

\item Taking the Boltzmann approximation of effective Yukawa couplings but diagonal number densities as the benchmark, the corrections result from the sum of the oscillation and interference terms. As is clear from the left panels of Fig.~\ref{fig:mixosc}, the corrections are large in the region $R\sim 1$, as one would expect, but the Boltzmann approximation agrees well with the total asymmetry elsewhere.

\item Taking the density matrix approximation of tree-level Yukawa couplings but off-diagonal number densities as the benchmark, the corrections result from the sum of the mixing and interference terms. Whilst the density matrix approximation (which captures the oscillation source) agrees well with the total asymmetry when the number density is equally shared amongst the two flavors, it is clear from the bottom right panel of Fig.~\ref{fig:mixosc} that keeping only the oscillation source underestimates the total asymmetry for smaller $R$ when there is a disparity between the occupancy of the two flavors. The latter result is consistent with the observed modification to the mixing source in Eq.~\eqref{eq:mixmod}.

\end{itemize}

We remark that care should be taken in extrapolating the above results to the strong washout regime in an expanding Universe and for a full phenomenological model. A more comprehensive discussion of this point and the implications of these observations for the treatment of flavor effects in low-scale leptogenesis scenarios, as well as the current status of various resummation procedures, is provided in Chapter~\cite{leptogenesis:A01} of this review.

\section{Strong washout approximation}\label{sec:strong}

In this section, we consider the strong washout regime of RL and summarize the main results of Ref.~\cite{Garbrecht:2014aga}.
This regime is phenomenologically interesting, and the discussion below is based on a realistic setup
taking the leading washout processes into account, as well as the expansion that drives the deviation
from equilibrium in this regime. We consider parametric regions where the 
mass splitting is much smaller than the average mass, i.e. $\Delta M \ll \bar{M}$, and the washout is strong, 
i.e. the right-handed neutrino relaxation rate exceeds the Hubble rate at $T=\bar{M}$.
The asymmetry is generated predominantly at temperatures below the Majorana neutrino mass,
such that we can use the non-relativistic approximations discussed in Sec.~\ref{sec:nrapp}.

In a radiation-dominated Universe, the scale factor is given by $a(\eta)=a_{\rm R} \eta$, where a convenient choice is $\eta=1/T$, which requires
$a_{\rm R}=M_{\rm Pl}\sqrt{45/(4 g_\ast \pi^3)}=T^2/H$, where $M_{\rm Pl}=G^{-1/2}$ is the Planck mass 
and $g_\ast$ is the effective number of degrees of freedom. Furthermore, it can easily be related to the parameter $z\equiv\bar{M}/T=\bar{M} \eta$,
which is often used in leptogenesis calculations. Reparametrizing Eq.~\eqref{eq:dn0hij} in terms of the parameter $z$ yields
\begin{subequations}
	\label{eq:dn0hijdz}
\begin{align}
\label{eq:dn0hijdz1}
	\bar{M} \frac{{\rm d}}{{\rm d}z}\delta n_{0h}\: +\: \frac{i a_{\rm R} z}{2 \bar{M}^2}\left[ M^2, \delta n_{0h}\right]
	\:+\: a_{\rm R} z \frac{g_w}{32 \pi}\left\{ {\rm Re}\,\lambda^\dagger \lambda , \delta n_{0h}\right\}
	\:+\: \bar{M} \frac{{\rm d}}{{\rm d}z}n^{{\rm eq}} \ &=\ 0\,,\\
\label{eq:dn0hijdz2}
	\bar{M} \frac{{\rm d}}{{\rm d}z}\delta \bar{n}_{0h}\: -\: \frac{i a_{\rm R} z}{2 \bar{M}^2}\left[ M^2, \delta \bar{n}_{0h}\right]
	\:+\: a_{\rm R} z \frac{g_w}{32 \pi}\left\{ {\rm Re}\,\lambda^\dagger \lambda , \delta \bar{n}_{0h} \right\}
	\:+\: \bar{M} \frac{{\rm d}}{{\rm d}z}n^{{\rm eq}} \ &=\ 0\,,
\end{align}
\end{subequations}
where the equilibrium number density is given by
\begin{align}
	n^{\rm eq} \ =\  \frac{z^2 \mathcal{K}_2(z)}{2 \pi^2}\,{\rm diag}(1,1) \ \approx\ \frac{z^{3/2} e^{-z}}{(2 \pi)^{3/2}}\,{\rm diag}(1,1)
	\label{neqdensity}
\end{align}
and $\mathcal{K}_{\nu}(z)$ are the modified Bessel functions of the second kind. In the fully-flavored approximation (for discussions of flavor effects in leptogenesis~\cite{Abada:2006fw, Nardi:2006fx, Abada:2006ea, Blanchet:2006be, Pascoli:2006ie, DeSimone:2006nrs}, see the acompanying Chapter~\cite{leptogenesis:A01}), any off-diagonal charge correlations in the SM charges are deleted by the lepton-flavor-violating
interactions mediated through the SM Yukawa-couplings (for a detailed explanation, see Ref.~\cite{Beneke:2010dz}).
The SM comoving charge densities are then governed by the following equation:
\begin{align}
	\label{eq:deltaLalphaalpha}
	-\,\bar{M} \frac{{\rm d}}{{\rm d} z} \Delta_{\alpha}\ &=\ g_w S_{\alpha \alpha}\: -\: W_{\alpha \alpha} n_{L \alpha}\: -\: \frac12 W_{\alpha \alpha} n_{\phi}\,,\\
	\nonumber
	&\equiv\ 4 \epsilon_{\alpha \alpha}(z) \bar{M} \frac{{\rm d} n^{\rm eq}}{{\rm d}z}\: -\: W_{\alpha \alpha} n_{L \alpha}\: -\: \frac12 W_{\alpha \alpha} n_{\phi}\,,
\end{align}
where  $n_\phi$ is the charge density in Higgs bosons. Equation~\eqref{eq:deltaLalphaalpha} has been written in terms of the asymmetries $\Delta_{\alpha} = n_B/3-n_{L \alpha}$, where $n_B$ is the baryon
number density, which are conserved by SM interactions. The washout matrix $W_{\alpha \beta}$ is given by (c.f.~Refs.~\cite{Beneke:2010dz,Blanchet:2011xq,Dev:2014laa})
\begin{align}
	W_{\alpha \beta} \ &=\ \sum_i \lambda_{\alpha_i} \lambda^*_{\beta i}
	\frac{3 a_{\rm R}}{(2\pi)^{3}}z^3 \mathcal{K}_{1}(z)\,\\
	&\approx\  \sum_i \lambda_{\alpha_i} \lambda^*_{\beta i}
	\frac{3 a_{\rm R}}{2^\frac72 \pi^\frac52}z^\frac52 e^{-z}\,.
	\label{def:washout}
\end{align}
Consistently with Eq.~\eqref{eq:deltaLalphaalpha}, we can define the time-dependent flavored effective decay asymmetry
\begin{align}
	\epsilon_{\alpha \beta}(z) \ &\equiv\
	g_w S_{\alpha \beta}(z) \left(4 \frac{{\rm d} n^{\rm eq}}{{\rm d}z} \right)^{-1} \bar{M}^{-1}\\
	&=\ \frac{1}{32 \pi} \frac{a_{\rm R} z }{\bar{M}} \sum_{i,j} \lambda_{\alpha i} \lambda^*_{\beta j}
	\left(\delta n_{0hij} - \delta \bar{n}_{0hij} \right) \left(\frac{{\rm d} n^{\rm eq}}{{\rm d}z} \right)^{-1}\,,
	\label{epsilonalphabeta:z}
\end{align}
with the clear physical interpretation as the asymmetry yield per sterile neutrino dropping out of equilibrium, where the factor of $4$ comes from the
two helicity states of the two heavy neutrinos.

In Ref.~\cite{Iso:2014afa}, it has been suggested that for large entries of $\lambda$, which correspond to a stronger washout, one may neglect the first
terms of Eq.~\eqref{eq:dn0hijdz1} and Eq.~\eqref{eq:dn0hijdz2}, i.e.~if the relaxation time for any of the heavy-neutrino states is shorter than the freeze-out time. Doing so leaves a system of algebraic equations that can be solved for the late-time limits of $\delta n_{0hij}(z)$ and $\delta \bar{n}_{0hij}(z)$. The solution for the off-diagonal correlations $\delta n_{0hij}=\delta\bar{n}_{0hij}^*$ $(i\neq j)$ is given by~\cite{Garbrecht:2014aga}
\begin{align}
	\delta n_{0hij}\ &=\ 
	\frac{{\rm Re}[\lambda^\dagger \lambda]_{ij}\left[(\lambda^\dagger\lambda)_{ii}+(\lambda^\dagger\lambda)_{jj}\right]}
		{[\lambda^\dagger \lambda]_{ii}[\lambda^\dagger \lambda]_{jj}}
		\\
	&\times \frac{
	\frac{\bar{M}^2}{8\pi}\left( [\lambda^\dagger \lambda]_{ii}+[\lambda^\dagger \lambda]_{jj} \right) 
	- i (M_i^2 - M_j^2)
	}{(M_i^2 - M_j^2)^2+A_{\rm eff}^2}\frac{\bar{M}^3}{a_{\rm R} z} \frac{{\rm d}}{{\rm d} z}n^{\rm eq}\,,
	\label{deltansol}
\end{align}
which leads to the late-time decay asymmetry
\begin{align}
	\epsilon_{\alpha \beta}^{\rm eff}\ =\
	-i (\lambda_{\alpha1}\lambda_{\beta 2}^*-\lambda_{\alpha 2}\lambda_{\beta 1}^*)\,
	\frac{{\rm Re}[\lambda^\dagger \lambda]_{12} [(\lambda^\dagger \lambda)_{11}+(\lambda^\dagger \lambda)_{22}]}
	{16 \pi (\lambda^\dagger \lambda)_{11}(\lambda^\dagger \lambda)_{22}}\,
		\frac{\bar{M}^2(M_2^2-M_1^2)}{(M_1^2-M_2^2)^2+A_{\rm eff}^2}\,,
	\label{eq:epsilonalphabetalatetime}
\end{align}
with the same effective regulator as obtained in the Wigner-space approach from Sec.~\ref{sec:weak}:
\begin{align}
	A_{\rm eff} \ =\  \frac{\bar{M}^2}{8\pi}[(\lambda^\dagger \lambda)_{11}+(\lambda^\dagger \lambda)_{22}]
	\left( 1 - \frac{[{\rm Re}(\lambda^\dagger \lambda)_{12}]^2}{(\lambda^\dagger \lambda)_{11}(\lambda^\dagger \lambda)_{22}} \right)^{1/2}\,.
\end{align}
As a final comparison, we can look at the ``unflavored'' decay asymmetry. By summing over the active flavors, we obtain
\begin{align}
	\epsilon^{\rm eff}\ =\ \sum_{\alpha} \epsilon_{\alpha \alpha}^{\rm eff} \ &=\ 
	\frac{{\rm Im}[(\lambda^\dagger \lambda)^2_{21}] [(\lambda^\dagger \lambda)_{11}+(\lambda^\dagger \lambda)_{22}]}
	{16 \pi (\lambda^\dagger \lambda)_{11}(\lambda^\dagger \lambda)_{22}}
		\frac{\bar{M}^2(M_2^2-M_1^2)}{(M_1^2-M_2^2)^2+A_{\rm eff}^2}\,,\\
		&=\frac{\epsilon_{11}^{{\rm eff}}+\epsilon_{22}^{{\rm eff}}}{2}\,,
	\label{eq:epsilonlatetime}
\end{align}
which corresponds to an average of the two diagonal decay asymmetries derived in the Wigner-space approach from Sec.~\ref{sec:weak}.

\subsection{Applicability of approximations}\label{sec:applicability}

As mentioned above, the key assumption for the strong washout approximation is that all of the elements of the density matrix $\delta n_{0h}$ have
reached their late-time limits. The precise way of quantifying this criterion is to compare the eigenvalues of the system of equations~\eqref{eq:dn0hij}
to the Hubble rate. At first, we make the assumption of only one lepton flavor. The late-time effective decay asymmetry then takes the form
\begin{align}
	\epsilon^{\rm eff} \ =\  \frac12\,\frac{X \sin(2 \varphi)}{X^2 + \sin^2(\varphi)}
	\label{epsilon:1flv}\,,
\end{align}
where we have introduced the phase $\varphi = \arg(\lambda_2/\lambda_1)$ between the Yukawa couplings and the dimensionless parameter
\begin{align}
	X\ =\ 8\pi\,\frac{M_1^2-M_2^2}{\bar{M}^2 \left( |\lambda_1|^2 + |\lambda_2|^2 \right)}\;,
	\label{xdef}
\end{align}
which can be interpreted as the ratio of the mass splitting and the mean of the decay widths.

Expressed in terms of the usual washout parameters~\cite{Buchmuller:2004nz} $K_i = |\lambda_i|^2 \bar{M}/(8 \pi H)|_{T=\bar{M}}$ and $\bar{K}= (K_1+K_2)/2$,
the smallest eigenvalue of the system of equations~\eqref{eq:dn0hijdz} is given by
\begin{align}
	\kappa \ =\  z \left[ \bar{K} - {\rm Re}
	\sqrt{K_1^2+K_2^2-2i(K_1^2-K_2^2) X - (K_1+K_2)^2 X^2 + 2 K_1 K_2 \cos 2\varphi}
	\right]\,.
\end{align}
If we restrict ourselves to the democratic case $|\lambda_1| = |\lambda_2|$, this simplifies to
\begin{align}
	\kappa \ =\  \bar{\kappa} \left[ 1 - \theta(\cos^2 \varphi - X^2 ) \sqrt{\cos^2 \varphi - X^2} \right]\,,
\end{align}
with $\bar{\kappa}= z \bar{K}$, where $\theta$ is the Heaviside step function. The condition that the slowest eigenmode is faster than the Hubble expansion rate is then satisfied if
$\kappa \gg 1$ around the time of freeze-out $z=z_f=O(10)$, allowing us to neglect the derivatives acting on $\delta n_{0hij}$ and $\delta \bar{n}_{0hij}$.
Therefore, for the approximation to be valid, the washout strength needs to satisfy
\begin{align}
	\bar{K}\ \gg\ (1/z_f)(\bar{\kappa}/\kappa)\,,
	\label{eq:washoutstr}
\end{align}
which allows us to use the ratio $\bar{\kappa}/\kappa$ to estimate the washout strength necessary for the applicability
of the strong washout approximation.

\begin{figure}[!t]
	\centering
	\includegraphics[width=0.7\textwidth]{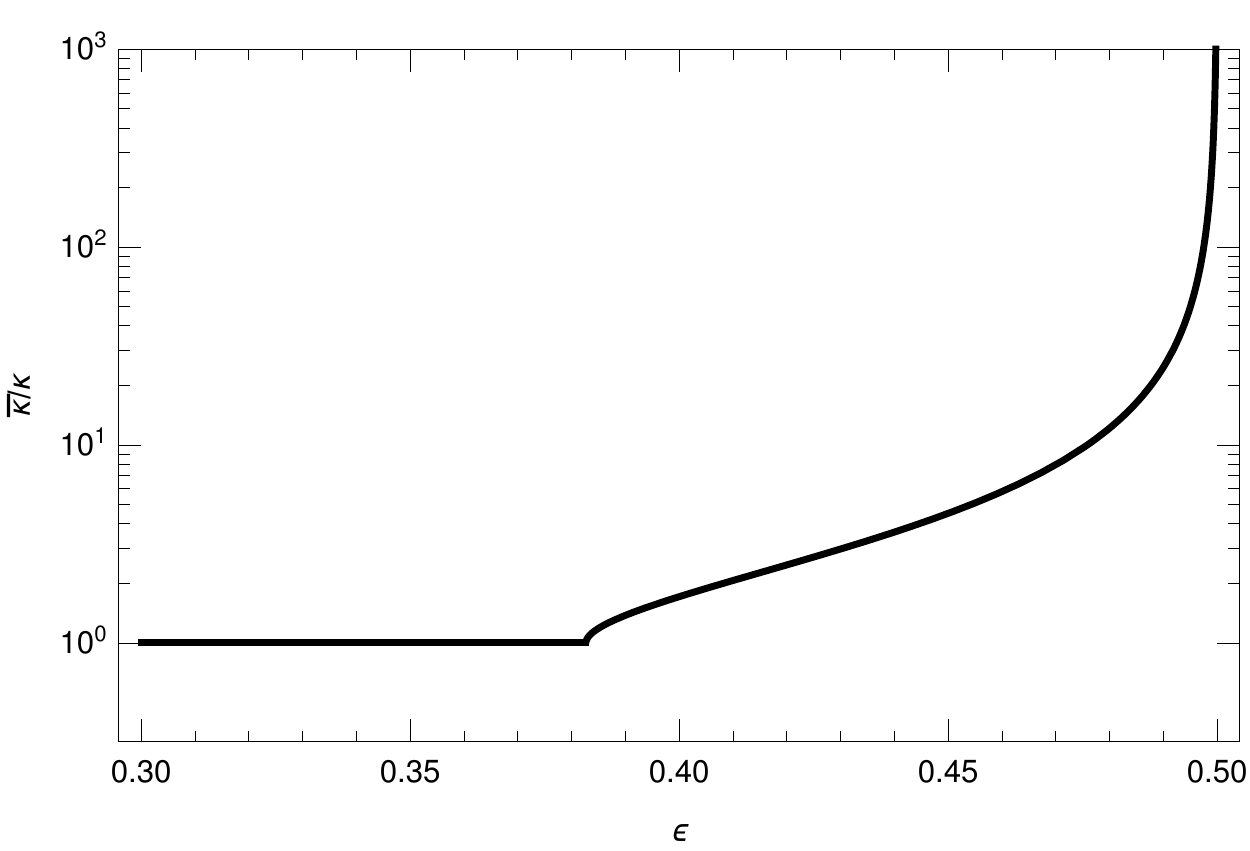}
	\caption{The relation between the decay asymmetry $\epsilon$ and the minimal ratio $\bar{\kappa}/\kappa$ of the smallest relaxation rate.
	For the derivatives of $\bar{\kappa}/\kappa$ to be negligible, the washout strength has to satisfy relation~\eqref{eq:washoutstr}.
	Note that, for $\epsilon \rightarrow 1/2$, it takes a longer time to build up the off-diagonal correlations, requiring a stronger washout
	for the approximations to be valid.
	}
	\label{fig:epskappa}
\end{figure}

We consider here the extreme cases, where the decay asymmetry reaches its maximal value. The angle $\varphi$ that maximizes the asymmetry is given by
\begin{align}
	\varphi_M \ =\ \arctan \frac{X}{\sqrt{1+X^2}}\,,
\end{align}
which leads to a decay asymmetry
\begin{align}
	|\epsilon| \ =\ \frac{1}{2\sqrt{1+X^2}}\;.
\end{align}
The effective decay asymmetry reaches its maximal value $|\epsilon| = 1/2$ for $X\rightarrow0$. It is interesting that, in this limit, the
CP-violating phase also vanishes as $\varphi_M\rightarrow 0$. However, the limit is not reachable in practice,
since the off-diagonal modes responsible for the CP asymmetry would require an infinite time to build up, because $\kappa/\bar{\kappa}\rightarrow 0$.
This implies that large decay asymmetries will be associated with smaller eigenvalues $\kappa$, which can easily be seen if we
express the relation between $\kappa/\bar{\kappa}$ and $\epsilon$ for $\varphi=\varphi_M$ as follows:
\begin{align}
\frac{\kappa}{\bar{\kappa}} \ =\  1\:-\: \theta\left[\epsilon^2 - \frac14\left( 2-\sqrt{2} \right)\right]
\sqrt{\frac{[\epsilon^2 - \frac14( 2-\sqrt{2} )][\epsilon^2 - \frac14( 2+\sqrt{2} )]}{\epsilon^2 (\epsilon^2-1/2)}}\,.
	\label{kkeps}
\end{align}
Note that this relation also gives the largest value of $\kappa/\bar{\kappa}$ for a fixed $\epsilon$, which is presented in Fig.~\ref{fig:epskappa}.

\begin{figure}[!t]
	\centering
	\includegraphics[width=0.7\textwidth]{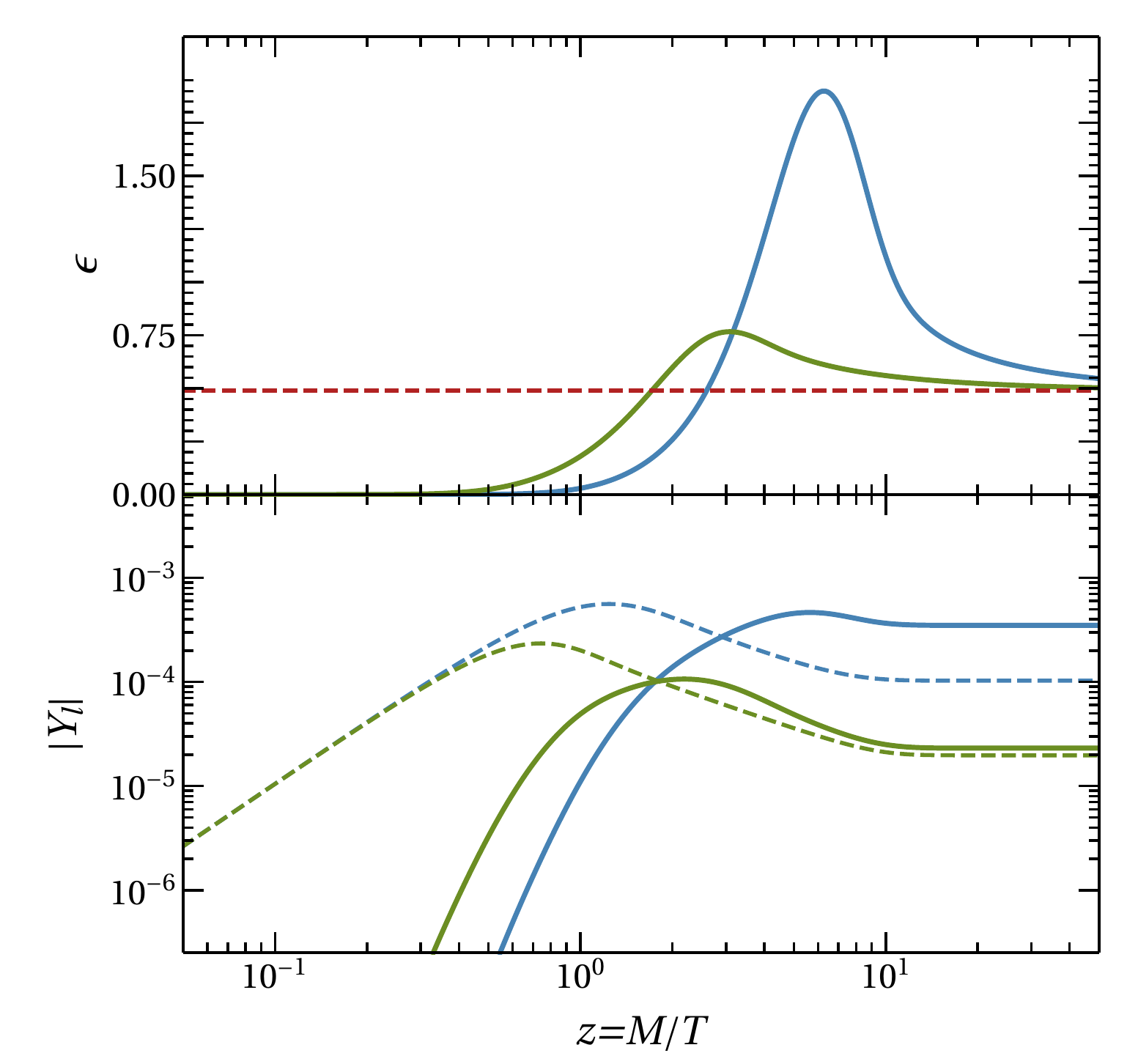}
	\caption{Upper panel: The time-dependent decay asymmetry $\epsilon(z)$ as defined in~\eqref{epsilonalphabeta:z}
	and its evolution towards the late-time limit $\epsilon=0.49$ (red, dotted)	for two values of the washout strength $\bar{K}=5$ (blue) and
	$\bar{K}=20$ (green).
	Lower panel: Evolution of the lepton asymmetry $|Y_L| = |n_L|/s$ normalized to the result obtained using the time-dependent decay asymmetry for different
	values of the washout strength $\bar{K}= 5$ (blue, solid), and $\bar{K}=20$ (green, solid), compared to the result obtained when using the late
	time limit (blue, dashed) and (green, dashed).
	}
	\label{fig:1flvstrongwashout}
\end{figure}

In Fig.~\ref{fig:1flvstrongwashout}, we show the comparison between the numerical solutions of Eq.~\eqref{eq:dn0hijdz} and \eref{eq:deltaLalphaalpha}
and the results obtained when using the late-time effective decay asymmetry for different values of the washout strength $\bar{K}$, with the weaker
one violating the criterion~\eqref{eq:washoutstr} and the other marginally complying with it.
One can see that the time-dependent effective decay asymmetry $\epsilon(z)$, as defined by Eq.~\eqref{epsilonalphabeta:z}, approaches its late-time limit
around $z_f=\mathcal{O}(10)$, where using the late-time limit becomes a good approximation. For simplicity, we neglected the charge in the Higgs field $n_\phi=0$ and take $n_L = \Delta$ (i.e.~$n_B=0$). Initially, using the late-time limit leads to quite a large discrepancy for both washout strengths. However,
around the freeze-out time, the discrepancy reduces to a factor four for the smaller washout, which does not satisfy relation~\eqref{eq:washoutstr}
and leads to a $20\%$ error for the larger washout.

We now consider a more realistic model with three lepton flavors. We consider the simple case with only two RH neutrinos.
The constraints from neutrino oscillation data are taken into account by using the Casas-Ibarra parametrization of the Yukawa couplings~\cite{Casas:2001sr}
\begin{align}
	\lambda \ \approx\  \frac{i\sqrt{2}}{v} U_{\nu} \sqrt{M_\nu^{\rm diag}} \mathcal{R} \sqrt{M_N^{\rm diag}}\;,
	\label{eq:CIparam}
\end{align}
where the diagonal mass matrix of the light neutrinos is given by $M_\nu^{\rm diag}$,
$U_\nu$ is the PMNS matrix with the best-fit parameters from the analysis in \cite{Esteban:2016qun}, $v=246\ {\rm GeV}$ is the expectation value of the Higgs field, $M_N^{\rm diag}$ is the diagonal mass matrix of the RH neutrinos and we have introduced the complex orthogonal matrices
\begin{align}
	\mathcal{R}^{\rm NO} \ =\
	\begin{pmatrix}
		0 & 0\\
		\cos \omega & \sin \omega \\
		-\xi \sin \omega & \xi \cos \omega
	\end{pmatrix}
	\,,\qquad
	\mathcal{R}^{\rm IO}\ =\
	\begin{pmatrix}
		\cos \omega & \sin \omega \\
		-\xi \sin \omega & \xi \cos \omega\\
		0 & 0
	\end{pmatrix}\,,
\label{eq:Rmatparam}
\end{align}
where $\rm NO$ and $\rm IO$ signify normal and inverted neutrino mass orderings, respectively.
We consider temperatures above $10^8\,{\rm GeV}$, where all second-generation but none of the first-generation Yukawa couplings are in equilibrium.
In this temperature regime, one can relate the number densities $n_L$ and $n_\phi$ to the asymmetries $\Delta$ via
\begin{align}
	n_L \ =\ A \Delta\,, \qquad n_\phi \ =\ C_\phi \Delta\,,
\end{align}
with
\begin{align}
	A\ =\ \frac{1}{1074}
	\begin{pmatrix}
	-906 & 120 & 120 \\
	75 & -688 & 28 \\
	75 & 28 & -688
	\end{pmatrix}
	\,,\qquad 
	C_\phi \ =\ -\,\frac{1}{179}
	\begin{pmatrix}
	37 & 52 & 52
	\end{pmatrix}\,.
\end{align}

Using the Casas-Ibarra parametrization, one can find a lower bound on the eigenvalues $\kappa/\bar{\kappa}$ to zeroth order in the mass splitting
$M_2-M_1$:
\begin{align}
	(\kappa/\bar{\kappa})^{\rm CI \,,NO} \ =\  \frac{m_2+m_3 \pm (m_3 - m_2) {\rm sech}(2 {\rm Im} \omega)}{m_2+m_3}\,,\\
	(\kappa/\bar{\kappa})^{\rm CI \,,IO}\ =\  \frac{m_1+m_2 \pm (m_2 - m_1) {\rm sech}(2 {\rm Im} \omega)}{m_1+m_2}\,,
\end{align}
where $m_1$, $m_2$ and $m_3$ are the active neutrino masses. Considering that the smallest ratio is $(\kappa/{\bar \kappa})\gtrsim 0.29$ for normal ordering (NO) and $(\kappa/{\bar \kappa})\gtrsim 0.99$ for inverted ordering (IO) [see criterion~\eqref{eq:washoutstr}],
for a phenomenological model with two RH neutrinos, it is a good approximation to neglect the
derivatives acting on $\delta n_{0hij}$ throughout the strong washout regime.
Note that, in the scenario with two RH neutrinos, the mean washout strength can be approximated by
\begin{align}
	\bar{K} \ =\  \left. \frac{\bar{M} {\rm tr}{\lambda^\dagger \lambda}}{16 \pi H}\right|_{T\,=\,{\bar M}} \approx
	\begin{cases}
		\mathcal{O}(30) \cosh(2 {\rm Im} \omega)\,,\quad \text{for NO\,,}\\
		\mathcal{O}(50) \cosh(2 {\rm Im} \omega)\,,\quad \text{for IO\,,}
	\end{cases}
\end{align}
which means that the washout is always strong in this scenario.

\begin{figure}[!t]
	\centering
	\includegraphics[width=0.7\textwidth]{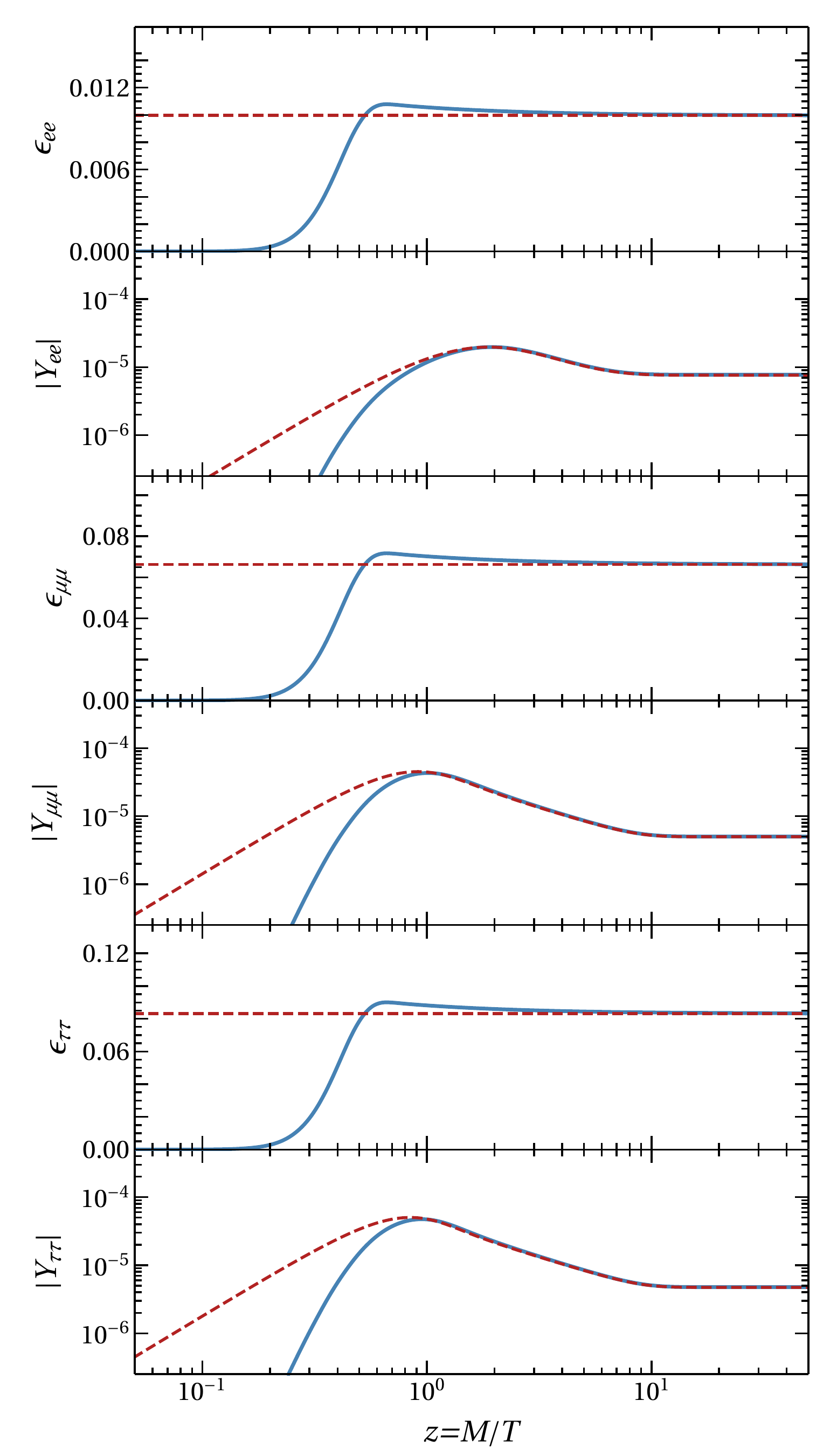}
	\caption{Comparison of the time-dependent decay asymmetries $\epsilon_{\alpha\alpha}(z)$ (solid) from Eq.~\eqref{epsilonalphabeta:z}
	and their late-time limits $\epsilon_{\alpha\alpha}$ (dashed) from Eq.~\eqref{eq:epsilonalphabetalatetime}.
	The parameters used are $\delta=0$, $\alpha=0$, $\omega = \pi/4 +0.2i$, $\Delta M /{\bar M}^2= -\,4\times 10^{-17}\,{\rm GeV}$.
	We also present the individual baryon-minus-lepton asymmetry yields $Y_{\alpha \alpha}= \Delta_{\alpha\alpha}/s$ obtained using
	the time-dependent decay asymmetry and the late-time limits.
	}
	\label{fig:swCIparam}
\end{figure}

In Fig.~\ref{fig:swCIparam}, we present the comparison between the results found when using the numerically-obtained time-dependent and
late-time effective decay asymmetries for all times prior to the freeze-out. 
Although the evolution of the asymmetries is different at early times, we get an $\mathcal{O}(1\%)$ agreement close to the freeze-out, as
expected from the eigenvalue arguments presented above.

\section{Model building and phenomenology of resonant leptogenesis}\label{sec:model}

In this section, we discuss some model building and phenomenological aspects of RL. As for the models, we review some simple RL scenarios, where the quasi-degeneracy of the heavy neutrino masses can be motivated {\it naturally}. This leads to a number of interesting experimental signatures of low-scale leptogenesis that can be tested at both energy and intensity frontiers. 

\subsection{Models of resonant leptogenesis}

As discussed above, a crucial requirement for RL models is the quasi-degeneracy of {\it at least} two  sterile neutrinos, which gives the resonant enhancement of the CP asymmetry and evades the Davidson-Ibarra bound \cite{Hamaguchi:2001gw, Davidson:2002qv}, thereby allowing one to lower the leptogenesis scale all the way down to the EW scale,\footnote{The sterile neutrino mass scale could be as low as GeV, but for mass scales well below the electroweak sphaleron freeze-out temperature, leptogenesis has to proceed either via sterile neutrino oscillations or Higgs decays (see the accompanying Chapter~\cite{leptogenesis:A02}), instead of sterile neutrino decay.} which can be tested in the foreseeable future. Additionally, such a low-scale mechanism has the advantage of being insensitive to assumptions about the thermal history of the Universe at much higher scales and could avoid dealing with potential problems of the creation of dangerous relics. Given these benefits, it is desirable to motivate the quasi-degeneracy of the RH neutrinos in the RL scenario from symmetry arguments, without resorting to fine-tuning. Below we discuss a few examples of such symmetry-protected RL scenarios. 

\subsubsection{Minimal testable resonant leptogenesis}\label{sec:minimal}

Let us first discuss models of RL in the minimal type I seesaw framework~\cite{Minkowski:1977sc, Mohapatra:1979ia, GellMann:1980vs, Yanagida:1979as} at or slightly above the electroweak scale. The light neutrino masses are generated via the standard type I seesaw mechanism 
\begin{equation}\label{eq:seesaw}
M_\nu \ \simeq \ -\frac{v^2}{2} \, \lambda \, M_N^{-1} \, \lambda^{T} \;,
\end{equation}
where $v = 246$ GeV is the electroweak vacuum expectation value (vev). For RH neutrinos in the $O(100)$ GeV range, the naive estimate for the size of the Yukawa couplings
\begin{equation}\label{eq:naive}
\lambda_{\rm naive} \ \sim \ \sqrt{\frac{M_\nu M_N}{v^2}}
\end{equation}
would give $\lambda_{\rm naive} \sim 10^{-7}-10^{-6}$, which is hard to test at current (or even proposed) experiments. In the following, we discuss  models of \emph{testable} RL, in which an appropriate symmetry protects the lightness of the neutrino masses, thus allowing values of (some of) the Yukawa couplings much larger than the naive estimate~\eqref{eq:naive}. Such symmetry is essentially a generalization of the ordinary lepton number $L$. In the case of two RH neutrinos, the same symmetry would also make them quasi-degenerate, thus providing a very minimal realization of testable RL~\cite{Blanchet:2009kk, Blanchet:2010kw}. However, it has been shown that, in the $L$-conserving limit, the asymmetry generated goes to zero at least at the same rate as the washout~\cite{Dev:2014laa}, and therefore obtaining successful leptogenesis and observable signatures in near-future experiments does not appear possible in this minimal  framework. 

The situation is drastically different with \emph{three} RH neutrinos~\cite{Pilaftsis:2004xx, Deppisch:2010fr}. In this case, the quasi-degeneracy between them, which allows for the desired resonant enhancement, can be guaranteed by imposing an approximate maximal $O(3)$ symmetry at some high scale $\mu_X$. Hence, at low scale the RH-neutrino mass matrix has the form~\cite{Pilaftsis:2004xx, Deppisch:2010fr,Dev:2014laa,Dev:2015wpa}
\begin{equation}\label{eq:MN}
M_N \ = \ m_N \bm{1} + \Delta M_N^{\rm RG} \ + \ \delta M_N \;,
\end{equation}
where $\Delta M_N^{\rm RG}$ is the contribution of the renormalization-group running from $\mu_X$ to the relevant scale $\mu \approx m_N$ and $ \delta M_N $ is a soft breaking term~\cite{Dev:2015wpa}, whose necessity stems from a no-go theorem discussed below in Sec.~\ref{sec:MFV}. In addition to this, the leptonic symmetry that allows to have testable RL is a $U(1)_L$, with charge assignment~\cite{Pilaftsis:2004xx,Pilaftsis:2005rv,Dev:2015wpa}: $L(N_1) = 0$, $L(N_\pm) = \pm 1$, $L(\ell_\alpha) = 1$, where $N_{\pm} = (N_2 \pm i N_3)/\sqrt{2}$ is the pseudo-Dirac combination.

Notice that, since $U(1)_L \sim O(2)_{N_{2,3}}$ breaks the original $O(3)$ symmetry, in a UV-completion of this model, one can imagine that the Yukawa couplings are switched on at some scale lower than $\mu_X$, for instance by some flavon fields acquiring a vev.

In this model, one can have, at the same time, successful leptogenesis and observables signatures at current and near-future experiments. The relevant phenomenology has been studied in Refs.~\cite{Dev:2014laa,Dev:2015wpa} and is reviewed in the accompanying Chapter~\cite{leptogenesis:A01}. 

\subsubsection{Minimal flavor violation and resonant leptogenesis} \label{sec:MFV}

The hypothesis of Minimal Flavor Violation (MFV) \cite{Buras:2000dm,DAmbrosio:2002vsn,Bobeth:2002ch} provides an elegant framework for naturally implementing the strong constraints from the non-observation of sizeable flavor-changing neutral current processes, at the same time having flavored new physics at the TeV scale. In the lepton sector, the realization of this idea is not unique, essentially because the mechanism for the generation of neutrino masses is unknown. In the so-called lepton MFV with extended field content~\cite{Cirigliano:2005ck}, one introduces a set of three RH neutrinos, and the light neutrino masses are generated via type I seesaw. In this case, the MFV hypothesis is that the only spurions that break the flavor symmetry are the Yukawa couplings $\lambda_{\alpha i}$, whereas the RH neutrino mass matrix is proportional to the identity at the scale $\mu_X$, and hence,
\begin{equation}\label{eq:MN_MFV}
M_N \ = \ m_N \bm{1} + \Delta M_N^{\rm RG}
\end{equation}
at the relevant scale $m_N$. Therefore, the absence of a soft-breaking term, as required by the MFV hypothesis, makes this different from that discussed above, cf.~Eq.~\eqref{eq:MN}. In particular, it has been shown~\cite{Dev:2015wpa,Pilaftsis:2015bja} that a no-go theorem prevents the generation of an asymmetry at $O(\lambda^4)$. This is because, at leading order in the renormalization-group running, one has 
\begin{equation}
 \Delta M_N^{\rm RG} \ \simeq \ - \,
 \frac{m_N}{8\pi^2} \ln\left(\frac{\mu_X}{m_N}\right)
 \mathrm{Re}\big(\lambda^\dag \lambda\big) \;.
\end{equation}
Therefore, the mass matrix in Eq.~\eqref{eq:MN_MFV} is diagonalized by a \emph{real} orthogonal matrix, which diagonalizes at the same time $\mathrm{Re}(\lambda^\dag \lambda)$. As a result, the CP asymmetry, proportional to $ \mathrm{Re}(\lambda^\dag \lambda)_{i j}$ ($j \neq i$) vanishes too.

In order to have a non-vanishing asymmetry, one has to go to the next order in the running~\cite{Pilaftsis:2015bja},\footnote{For a related discussion in the supersymmetric context, see Refs.~\cite{Babu:2008kp, Achelashvili:2017nqp}.} thus obtaining effectively a $O(\lambda^6)$ asymmetry. However, this implies that the Yukawa couplings for successful leptogenesis need to be larger than in the standard $O(\lambda^4)$ case and, by virtue of the seesaw relation~\eqref{eq:seesaw}, the RH neutrinos need to be significantly heavier than the electroweak scale, even if one is in the resonant regime automatically by virtue of the MFV hypothesis. One obtains the very stringent bound~\cite{Pilaftsis:2015bja} $m_N \gtrapprox 10^{12}$ GeV, significantly higher than previous results in the literature. However, in the context of the MFV approach, this high value of $m_N$ implies observable effects in LFV experiments, as discussed in the accompanying Chapter~\cite{leptogenesis:A01}. This is because, in the spirit of the MFV hypothesis, a flavor-violating new-physics scale $\Lambda_{\rm LFV} \sim O(\text{TeV})$ is assumed, and a large hierarchy between the lepton-\emph{number} violating scale $\Lambda_{\rm LNV} = m_N > 10^{12}$ GeV and $\Lambda_{\rm LFV}$ significantly boosts the LFV observables~\cite{Cirigliano:2005ck}.

\subsubsection{Inverse seesaw}

This is a variant of the type I seesaw, where two sets of SM-singlet fermions with opposite lepton numbers are added to the particle content of the SM~\cite{Mohapatra:1986bd}. This is a technically natural realization of the seesaw, which allows large active-sterile mixing without resorting to fine-tuning. Lepton number is approximately conserved in this case. The main difference between inverse and type I seesaw is that the light neutrino mass in the former case is {\it directly} proportional to the small lepton number breaking, and this freedom allows one to relax the constraints from neutrino oscillation data on the sterile neutrino mass scale.  

As far as resonant leptogenesis is concerned, the quasi-degeneracy of the RH neutrinos is naturally realized in the inverse seesaw setup, with the mass splitting proportional to the small LNV parameter in the theory~\cite{Blanchet:2009kk, Blanchet:2010kw}. In fact, both the lepton asymmetry and the LNV washout effects go to zero in the $L$-conserving limit. However, it has been shown that, in the $L$-conserving limit, the asymmetry generated goes to zero at least at the same rate as the washout~\cite{Dev:2014laa}, and therefore obtaining successful leptogenesis and observable signatures in near-future experiments does not appear possible in the minimal inverse seesaw framework.\footnote{This is in contrast with the past claims~\cite{Blanchet:2009kk, Blanchet:2010kw}, which used a regulator for the $CP$ asymmetry that has a pathological behavior in the $L$-conserving limit, thereby overestimating the lepton asymmetry by orders of magnitude.} One possible work-around is to further enlarge the fermion sector~\cite{Aoki:2015owa} or to go to the generalized inverse seesaw case with an appropriate flavor structure~\cite{Dev:2012sg}.

\subsubsection{Froggatt-Nielsen mechanism}

Another rather generic scenario that realizes the resonant leptogenesis idea is based on the Froggatt-Nielsen (FN) mechanism~\cite{Froggatt:1978nt}. In this case, one introduces two FN fields, $\Sigma$ and $\overline\Sigma$, with opposite $U(1)_{\rm FN}$ charges, and also makes two RH neutrinos oppositely charged under this symmetry, while all other fields are singlets under $U(1)_{\rm FN}$. For an appropriate choice of $\langle\Sigma\rangle$ and $\langle\overline\Sigma\rangle$, one could get mass splittings naturally of order of the decay width and realize a TeV-scale resonant leptogenesis~\cite{Pilaftsis:2003gt}. See also Refs.~\cite{Asaka:1999jb, AristizabalSierra:2007ur, Kamikado:2008jx}.

\subsubsection{Soft supersymmetry breaking}

This class of models is motivated from a symmetry to forbid the $\mu H_uH_d$ superpotential term, which only arises due to an intermediate-scale SUSY breaking term within a higher-dimensional, Planck-suppressed operator. In a RH-neutrino extended MSSM, such a symmetry can also suppress the masses and interactions of the RH neutrinos, and naturally lead to resonant leptogenesis~\cite{Hambye:2004jf, Grossman:2004dz, Boubekeur:2004ez, Fong:2010qh, Fong:2011yx}. For more details, we refer to the accompanying Chapter~\cite{leptogenesis:A05}.

\subsubsection{Flavor symmetry} 

Many models based on discrete flavor $G_f$ and CP symmetries can naturally realize a degenerate heavy-neutrino spectrum, which is then softly broken to realize RL. Examples include $G_f=\Delta(3n^2)$~\cite{Luhn:2007uq} or $\Delta(6n^2)$~\cite{Escobar:2008vc} (with $n$ even, $3 \nmid n$, $4 \nmid n$), where the LH lepton doublets transform in a ${\bf 3}$, the RH neutrinos in a ${\bf 3}'$ and RH charged leptons in a ${\bf 1}$~\cite{Hagedorn:2016lva, Claudia}; see Refs.~\cite{Xing:2006ms, Branco:2009by, Bertuzzo:2009im,Sierra:2011vk} for other examples in which the discrete symmetries (without CP) were used in order to explain the pattern of lepton mixing and to realize resonant leptogenesis. 
A detailed discussion of flavor symmetries in the context of leptogenesis can be found in the accompanying Chapter~\cite{leptogenesis:A06}.

\subsubsection{Extended gauge sector}

Apart from models that guarantee the quasi-degeneracy of the RH neutrinos, it is interesting to consider the embedding into frameworks where the very existence of (typically 3) RH neutrinos can be explained, either by unification or anomaly cancellation. Typically, the additional ingredients present in this framework, if their masses are in the TeV range or less, may significantly alter the phenomenology, and it is therefore important to take them into account.\footnote{Similar leptogenesis constraints can also be derived on the new scalar masses in both $U(1)$ and left-right models~\cite{Dev:2017future}. Thus, a future observation of such new Higgs bosons could effectively test the RL scenario.} 

In this respect, a prime example is given by left-right symmetric models, where the presence of three RH neutrinos is required by unification with the RH charged leptons, together with the left-right symmetry. In this case, there exist processes where the RH gauge boson $W_R$ is exchanged in scattering processes involving only one RH neutrino. These processes do not efficiently decouple by Boltzmann suppression~\cite{Frere:2008ct, Dev:2014iva, Dev:2015vra} (unlike the minimal $U(1)_{B-L}$ case, where the $Z'$-mediated scatterings are doubly Boltzmann suppressed~\cite{Blanchet:2009bu}) and thus RH gauge interactions keep $N$ at equilibrium very efficiently until temperatures much smaller than $m_N$. As a result, leptogenesis becomes impossible unless the mass of the $W_R$ is higher than $\sim 10$ TeV~\cite{Frere:2008ct,Dev:2015vra}, thus making the potential observation of the $W_R$ at the LHC incompatible with leptogenesis. For more details, see Chapter~\cite{leptogenesis:A05}.

\begin{figure}[!t]
\centering
\includegraphics[width=0.5\textwidth]{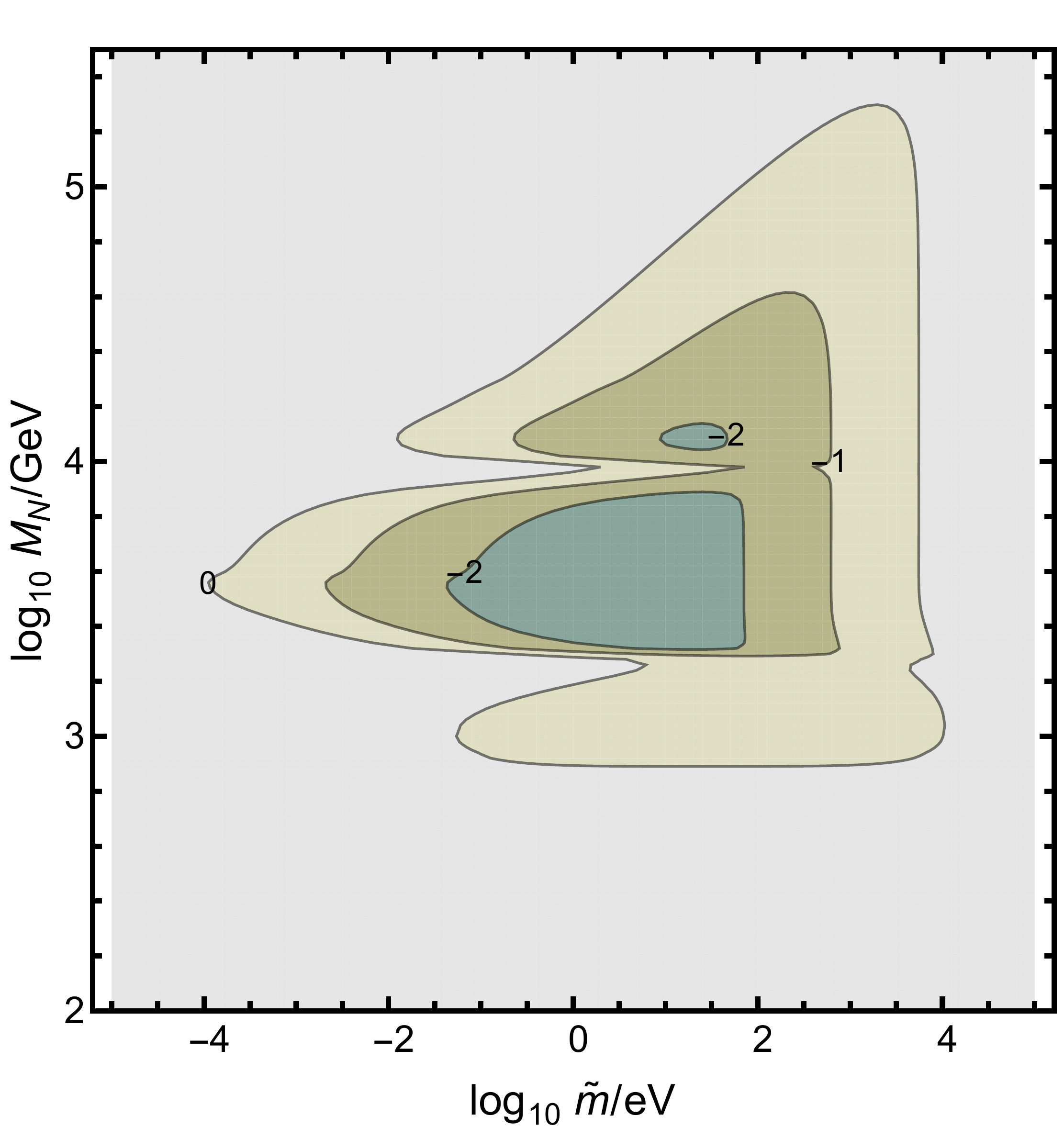}
\caption{The logarithm base 10 of the CP asymmetry  in the decay $N \to \ell \phi$ needed for successful leptogenesis, as a function of the mass of the RH neutrinos and the effective light neutrino mass parameter $\tilde m = v^2 (\lambda \lambda^\dag)/m_N$, in the one-flavor approximation. For illustrative purposes, we have chosen the mass of the $Z'$ to be 4 TeV, the one of  the $B\!-\!L$ breaking scalar to be 20 TeV and the $B\!-\!L$ gauge coupling equal to $0.5$.
\label{fig:Zprime}}
\end{figure}

Another important example where three RH neutrinos are needed, this time because of anomaly cancellation, is in the case of an Abelian $B\!-\!L$ symmetry \cite{Buchmuller:1991ce, Basso:2008iv}. In this case (and more generally in the presence of a \emph{neutral} extra gauge boson), gauge processes involve \emph{two} RH neutrinos. Therefore, differently from above, the decoupling of these proceeds via a Boltzmann suppression~\cite{Frere:2008ct}. As a result, even for $O(1)$ gauge couplings, the RH neutrinos may exit equilibrium at a temperature not too far below their mass, thus making leptogenesis possible, although typically very constrained~\cite{Racker:2008hp,Blanchet:2009bu,Heeck:2016oda}. For instance, following the analysis of Ref.~\cite{Heeck:2016oda}, in Fig.~\ref{fig:Zprime}, we plot the region of successful leptogenesis, for an illustrative choice of parameters. Generic textures in the seesaw relation, without large cancellations between different entries, correspond to values of the parameter $\widetilde{m} \approx 50 $~meV. For the choice of parameters as in Fig.~\ref{fig:Zprime}, leptogenesis is then possible only for $200\, \text{GeV} \lesssim m_N \lesssim 800\,\text{GeV}$ and with a CP asymmetry close to its possible maximal value of 1. Allowing for cancellations in the seesaw relation, possibly due to a symmetry, as e.g.~in Sec.~\ref{sec:minimal}, clearly enlarges the parameter space.
 
For a heavy $Z'$ in the TeV range, the phenomenology has been studied in Ref.~\cite{Blanchet:2009bu} and more recently in Ref.~\cite{Heeck:2016oda}, in the latter case also taking into account the important effects of the scalar that breaks the $B\!-\!L$ symmetry. The general picture that emerges is that, although the presence of a $Z'$ with $m_{Z'} > 2 m_N$ improves significantly the discovery prospects of $N$ at colliders, the presence of a $Z'$ in the TeV range, with large gauge coupling, makes leptogenesis very constrained. For light $Z'$ in the GeV range or less, a number of near-future experiments make this regime particularly interesting. In this case, as studied in Ref.~\cite{Heeck:2016oda}, the prospects of having successful leptogenesis in the region of parameter space tested by these experiments are quite good. Successful leptogenesis provides interesting constraints on the mass of the RH neutrinos and the $B\!-\!L$ gauge coupling precisely in the region that is phenomenologically accessible.

Some more phenomenological aspects of models realizing resonant leptogensis, beyond the type I seesaw, are discussed in the accompanying Chapter~\cite{leptogenesis:A05}. For more details, see e.g.~Ref.~\cite{Hambye:2012fh}.

\section{Conclusion}\label{sec:conclusions}

The scenario of resonant leptogenesis is widely studied due to its phenomenological relevance. A profound theoretical underpinning
of its dynamics is an important ingredient to assess potential signatures in upcoming experimental campaigns, including the future 
LHC programme, searches for lepton flavor violation, lepton number violation and neutrinoless double beta decay, and future lepton and/or hadron colliders.

In this article, we reviewed the status of theoretical descriptions that capture the saturation of resonant enhancement and provided
an overview of selected models realizing a quasi-degenerate mass spectrum in conjunction with a sizeable CP asymmetry.
When the mass difference $\Delta M$ is of order the decay width $\Gamma_i$ of the Majorana neutrinos, it is necessary to employ a framework for obtaining quantum Boltzmann equations that accounts fully for flavor coherence in the sterile sector. We discussed three complementary approaches based on the closed-time-path approach that address
this regime, dubbed two-time (or Kadanoff-Baym), Wigner, and two-momentum/interaction-picture descriptions.
The three approaches are complementary in that they allow one to reduce the full non-equilibrium evolution equations for the
lepton asymmetry and the Majorana neutrino correlation functions to a tractable
set of equations based on different underlying assumptions. 

All approaches confirm the basic mechanism of resonant enhancement
and agree with the usual Boltzmann description for $\bar M \gg \Delta M \gg \bar M(\lambda^\dag\lambda)_{ij}/(8\pi)$ and  $T\ll \bar M$,
where $\bar M$ is the average mass and $T$ the relevant temperature during the production of the asymmetry.
For smaller mass splittings, the final asymmetry can, in general, be determined only by solving (numerically) a set of equations
that takes flavor covariance into account. We considered various simplified limits that allow to compare the different approaches.
For example, the two-time equations can be solved without further approximations in a static setup, where they agree with the
Wigner approach within its expected regime of validity (in the narrow-width limit and for $\bar M \gg \Delta M$). When approaching a hierarchical
spectrum, the Wigner approach discussed here breaks down, while the usual Boltzmann treatment is expected to become valid. This can
be checked with the two-time approach, which interpolates between the Wigner result for $\bar M \gg \Delta M$ and the Boltzmann
result for $\Delta M \gg \bar M(\lambda^\dag\lambda)_{ij}/(8\pi)$. The two-momentum/interaction-picture approach in addition
provides the possibility to disentangle contributions to the final asymmetry due to coherent transitions and decays.
It provides the basis for a large number of phenomenological studies and can also be extended in a straightfoward way
to account for active- in addition to sterile-flavor effects.
We reviewed an application of this approach to a scalar toy model and commented on the relation to the two-time approach
in this case. It will be interesting to extend this comparison to the type I seesaw model in the future.

Finally, it is interesting to note that the full dynamics can be captured by a simplified
approach under certain circumstances, where it is possible to express the final asymmetry in terms of the 
result that would be expected based on a Boltzmann treatment. Specifically, this is possible in the particular weak washout scenario considered in Sec.~\ref{sec:weak},
provided the flavor-coherence vanishes initially (cf.~e.g.~Eq.~\eqref{eq:cpeffweak} for $i=j$),
as well as in the well-known strong washout regime, discussed in Sec.~\ref{sec:strong} (c.f.~Eq.~\eqref{eq:epsilonlatetime}). 
Intriguingly, both correspond to an effective decay asymmetry of the form given in Eq.~(\ref{eq:cpA}) (or Eq.~\eqref{eq:epsilonalphabetalatetime} when
not summing over active flavors) with an effective regulator $A$ given by
\begin{equation}
  A_{\rm eff}\  \equiv\  \frac{\bar M^2}{8\pi}\left((\lambda^\dag\lambda)_{11}+(\lambda^\dag\lambda)_{22}\right)\left(1-\frac{[\mbox{Re}(\lambda^\dag\lambda)_{12}]^2}{(\lambda^\dag\lambda)_{11}(\lambda^\dag\lambda)_{22}}\right)^{1/2}\;.
\end{equation}
Nevertheless, it is important to take into account the limitations of the applicability of these simplified expressions  (cf.~Sec.~\ref{sec:applicability} for a quantitative discussion in the strong washout regime) in practical applications.

\section*{Acknowledgements}

We thank Laura Covi, Bj\"orn Garbrecht and Emiliano Molinaro for useful comments. The work of PM is supported by STFC Grant No. ST/L000393/1 and a Leverhulme Trust Research Leadership Award.  We gratefully acknowledge the hospitality of the Munich Institute for Astro- and Particle Physics (MIAPP) of the DFG cluster of excellence
``Origin and Structure of the Universe'', where this work has been initiated. The work of DT is supported by a ULB postdoctoral fellowship and the Belgian Federal Science Policy (IAP P7/37)

\bibliographystyle{ws-rv-van-mod2}
\bibliography{reslg_v2}

\end{document}